\documentstyle[12pt]{article}

\def\be{\begin{equation}}
\def\ee{\end{equation}}
\def\bea{\begin{eqnarray}}
\def\eea{\end{eqnarray}}

\def\br{\begin{thebibliography}{abc}}
\def\er{\end{thebibliography}}

\def\inbar{\,\vrule height1.5ex width.4pt depth0pt}
\def\IB{\relax{\rm I\kern-.18em B}}
\def\IC{\relax\hbox{$\inbar\kern-.3em{\rm C}$}}
\def\ID{\relax{\rm I\kern-.18em D}}
\def\IE{\relax{\rm I\kern-.18em E}}
\def\IF{\relax{\rm I\kern-.18em F}}
\def\IG{\relax\hbox{$\inbar\kern-.3em{\rm G}$}}
\def\IH{\relax{\rm I\kern-.18em H}}
\def\II{\relax{\rm I\kern-.18em I}}
\def\IK{\relax{\rm I\kern-.18em K}}
\def\IL{\relax{\rm I\kern-.18em L}}
\def\IM{\relax{\rm I\kern-.18em M}}
\def\IN{\relax{\rm I\kern-.18em N}}
\def\IO{\relax\hbox{$\inbar\kern-.3em{\rm O}$}}
\def\IP{\relax{\rm I\kern-.18em P}}
\def\IQ{\relax\hbox{$\inbar\kern-.3em{\rm Q}$}}
\def\IR{\relax{\rm I\kern-.18em R}}
\font\cmss=cmss10 \font\cmsss=cmss10 at 7pt
\def\IZ{\relax\ifmmode\mathchoice
{\hbox{\cmss Z\kern-.4em Z}}{\hbox{\cmss Z\kern-.4em Z}}
{\lower.9pt\hbox{\cmsss Z\kern-.4em Z}}
{\lower1.2pt\hbox{\cmsss Z\kern-.4em Z}}\else{\cmss Z\kern-.4em Z}\fi}
\def\IGa{\relax\hbox{${\rm I}\kern-.18em\Gamma$}}
\def\IPi{\relax\hbox{${\rm I}\kern-.18em\Pi$}}
\def\ITh{\relax\hbox{$\inbar\kern-.3em\Theta$}}
\def\IOm{\relax\hbox{$\inbar\kern-3.00pt\Omega$}}

\def\rt{\rightarrow}

\def\Hat#1{\rlap{\kern.10em$\widehat{\phantom G}$}#1}
\def\HAt#1{\rlap{\kern.05em$\widehat{\phantom G}$}#1}

\def\czp#1{\rlap{\kern.1em$\widehat{\phantom{G\vrule height.8em}}$}#1{}}
\def\Czp#1{\rlap{\kern.05em$\widehat{\phantom{G\vrule height.8em}}$}#1{}}

\def\semidirect{\mathbin
               {\hbox
               {\hskip 3pt \vrule height 5.2pt depth -.2pt width .55pt 
                \hskip-1.5pt$\times$}}}
\newcommand{\sect}[1]{\setcounter{equation}{0}\section{#1}}

\def\sxn#1{\bigskip\medskip \sect{#1} \smallskip
                                                 }

\begin{document}
\input epsf.tex
\thispagestyle{empty}
\setcounter{page}{0}

\begin{flushright}
SU-4240-698\\
\end{flushright}

\vspace{1cm}

\begin{center}
{\LARGE Quantum Topology Change  in $(2 + 1)d$ \\}
\vspace{5mm}
\vspace{2cm}
{A.P. Balachandran$^1$, E. Batista$^2$,  I.P. Costa e Silva$^3$ and 
P. Teotonio-Sobrinho$^3$}\\

\vspace{.50cm}
$^1${\it Department of Physics,Syracuse University \\
Syracuse, NY 13244-1130, USA}\\
\vspace{.50cm}
$^2${\it Universidade Federal de Santa Catarina, Centro de F\'isica e
  Matem\'atica,\\ 
Dep. MTM, CEP 88.010-970, Florian\'opolis, SC, Brazil}\\ 
\vspace{.50cm}
$^3${\it Universidade de S\~ao Paulo, Instituto de F\'isica-DFMA \\
Caixa Postal 66318, 05315-970, S\~ao Paulo, SP, Brazil}\\
\vspace{2mm}
\vspace{.2cm}

\end{center}

\begin{abstract}
The topology of orientable $(2+1)$ spacetimes can be captured by certain
lumps of non-trivial topology called topological geons. They are the
topological analogues of conventional solitons. We give a
description of topological geons where the degrees of freedom related
to topology are separated from the complete theory that contain
metric (dynamical) degrees of freedom. The formalism also allows us to
investigate processes of quantum topology change. They correspond to
creation and annihilation of quantum geons. Selection rules for such
processes are derived.

\end{abstract}

\newpage

\sxn{Introduction}\label{S1}

It is very common to make the 
reasonable assumption that the topology of space-time is fixed. We
assume that space-time is a manifold of the form
$\Sigma \times {\IR}$, and that for each \mbox{time $t$},
we have a space-like surface that is always homeomorphic to a given 
$\Sigma $. However, when (quantum) gravity is taken into
account, the very geometry of space becomes a degree of freedom, and
one can conceive the possibility that $\Sigma $ 
changes in the course of time \cite{Wheeler}. Such a process is called
topology change. Creation of baby universes, production of
topological defects (cosmic strings, domain walls), and changes in genus
(production of wormholes and topological geons) are examples 
of topology change. Each of them have received
some attention in the literature. 
Several authors have investigated topology change  
within the context of both 
classical and quantum gravity \cite{QG}. It is
interesting to notice that in the usual canonical approach to gravity,
only 
the metric of the spatial manifold $\Sigma $ appears as a
degree of freedom and receives a quantum treatment. The topology of 
$\Sigma $ in its turn is implicitly treated as a 
classical entity. There are, of course, other approaches to quantum
gravity such as  string theory \cite{string} and Euclidean quantum
gravity \cite{EGr} where topology may
appear as an entity of a quantum nature via a sum over topologies.

It would be desirable to have a formalism where topology
can in a certain sense be canonically quantized and if possible
separated from degrees of freedom coming from metric and other fields. 
In spite of the fact that topology change has been
inspired by quantum gravity, it has been demonstrated  in \cite{TC}
that it can happen in ordinary quantum mechanics. In this approach,
metric is not dynamical, 
but degrees of freedom related to topology are quantized. The notion of a 
space with a well defined topology appears only as a classical limit.
(See also \cite{marolf} for related ideas). The views we would like
to present in this paper are similar, to a certain extent,  
to the ones in \cite{TC}. In our approach, variables related to
topology are separated from other degrees of freedom and then quantized. 

The topology of space is well captured by soliton-like 
excitations of $\Sigma $ called topological geons. 
They can be thought of as lumps of 
nontrivial topology. 
For example, in \mbox{$(2+1)d$}, the
topology of an orientable, closed surface $\Sigma $ is determined by the
number of connected components of $\Sigma $ and by the number of of
handles on each connected component. Each handle corresponds to
a topological geon, i.e., a localized lump of nontrivial topology. 
It is well known that these
solitons have particle like properties such as spin and statistics.
However unlike ordinary particles they can violate the  spin-statistics 
relation \cite{erice,spin-stat}. It has been suggested
\cite{spin-stat,where,DS} that the standard spin-statistics
relation can be recovered if one considers processes where geons are
(possibly pairwise) created and annihilated, but this necessarily
implies a change of the topology of $\Sigma $. In other words, one may
have to consider topology change in order to have a spin-statistics
theorem for geons \cite{where,DS}. 

The Euclidean path integral approach can in some sense be carried
out in low dimensions \cite{sum}, but it represents a formidable task
in the case of  
a $(3+1)d$ theory. It would be nice to stay closer to a ``canonical''
quantization, even though topology change and the canonical approach
appear to be incompatible. One may search for alternative
descriptions of topological properties using algebraic tools, very
much in the spirit of quantum invariants of knot theory. The polynomial 
invariants of knots can be obtained by both  field
theoretic and algebraic methods. 
In the field theoretical approach, it is well known
that Jones polynomials are obtained by means of functional 
integrals of  Chern-Simons theory  \cite{witten}. In the
algebraic approach, one obtains invariants by
representations of the braid group \cite{jones, kauffman}, 
or via Hopf algebras \cite{kassel,turaev}.
We  will try in this paper to give an algebraic description of 
quantum geons, rather than a field theoretical one. We will present 
a theory of quantized topological geons where topology change is a
quantum transition. We will only analyze the case of orientable  geons
in $(2+1)d$ were
handles are the only possible ``particles''. A generalization to
include nonorientable geons will be presented elsewhere.

Let us consider a manifold $M$ and some generic field theory 
(possibly with  gauge and Higgs fields) 
interacting with gravity. It is reasonable to expect that if we 
could quantize such a complex theory, its observables would 
give us information on the geometry and 
topology of $M$. The main point is that one does not need to consider the
full theory to get some topological information. It is possible
that, in a certain low energy (large distance) limit,  there would be
a certain set of  observables encoding the topological data. 
We know examples where this is precisely the case. In general, the 
low energy (large distance) limit of a field theory is not able to probe
details of the short distance physics, but it  can isolate degrees of
freedom related to
topology. We may give as an example the low energy limit of $N=2$
Super Yang-Mills,
known as the Seiberg-Witten theory \cite{SW}. We also have examples of
more drastic reduction where a field theory in the vacuum state 
becomes purely topological \cite{vacuum}. Inspired by these facts 
we will identify the degrees of freedom, 
or the algebra ${\cal A}^{(n)}$ of ``observables'',  
capable of describing $n$ topological geons in $(2+1)d$. Actually, we
will argue later in this paper that the operators in this algebra are
not really observables in the strict sense. Rather, it is what is
called \cite{field} a {\it field algebra}. 
We say that ${\cal A}^{(1)}$ 
describes a single
geon  in the same way that the algebra of angular momentum describes a
single spinning particle. In this framework what we mean by quantizing
the system is nothing but finding  irreducible representations
of ${\cal A}^{(1)}$. As in the case of the algebra of angular momentum,
different irreducible representations have to be thought of as
different {\it particles}.  
For the moment, we will not be concerned with dynamical aspects. 
We would like to
concentrate on the quantization itself and leave the dynamics to be
fixed by the particular model one wants to consider.  

An intuitive way of understanding the algebra ${\cal A}^{(1)}$ 
for a topological geon comes from considering a gauge
theory with
gauge group $G$ in two space dimensions spontaneously broken to a
discrete group $H$. For simplicity we will assume that $H$ is
finite. As an immediate consequence it follows that 
the gauge connection (at far distances) is locally flat. In other
words, homotopic loops $\gamma $ and $\gamma '$ produce the same 
parallel transport (holonomy). The set of
independent holonomies are therefore parametrized by elements 
$[\gamma ]$ in the fundamental group $\pi _1(\Sigma )$.  
It is quite clear that such quantities are enough
to detect the presence of a handle. The phase space we are 
interested in contains only topological degrees of freedom. Therefore 
such holonomies can be thought of as playing the role of position variables. 
We also have to take into account the 
diffeomorphisms that are able to change $[\gamma ]$. They will be
somewhat the analogues of translations. 
It is clear that the connected component of the group of
diffeomorphisms, the so-called small diffeos, cannot
change the homotopy class of $\gamma $. 
To change the homotopy class of
a curve $\gamma $ one needs to act with the so-called large
diffeomorphisms. 
Therefore the analogues of
translations have to be parametrized by the large diffeos modulo the
small diffeos. This is exactly the mapping class group
$M_\Sigma $. Also, we must take into account an action of the group
$H$, changing the holonomies by a conjugation. This action, as we will
discuss later, corresponds physically to ``encircling flux sources at
infinity''. These three sets of
quantities will
comprise our algebra ${\cal A}^{(1)}$. Contrary to what happens in field
theory or even in quantum mechanics, we find that ${\cal A}^{(1)}$ is finite 
dimensional. This will be important to avoid technical problems of
various kinds. The algebra ${\cal A}^{(1)}$ contains the analogue of  
positions and translations and can be thought of as
a discrete Weyl algebra. There seems to be no great obstacle to generalize our
results also to the case where $H$ is a Lie group \cite{gen}.

Our algebraic description of geons is analogous to what has been developed
for  $2d$ non-abelian vortices by the Amsterdam group
\cite{prop}. These ideas have been further developed by some of us and
coworkers and applied to rings in $(3+1)d$. Their results will not be
discussed here since a complete account will be reported in \cite{syr}.

The algebra encountered by \cite{prop} was a special type of Hopf
algebra, namely the Drin'feld double of
a discrete group \cite{kassel}. In our case, however, 
the algebra ${\cal A}^{(1)}$ is not Hopf, but it has  a Drin'feld double
as a subalgebra. For a pair of geons we find that the corresponding
algebra ${\cal A}^{(2)}$ is 
closely related to the tensor product 
${\cal A}^{(1)}\otimes {\cal A}^{(1)}$ of
single geon algebras. This fact allows us to determine
the appropriate algebra ${\cal A}^{(n)}$ for an arbitrary number $n$ of
geons.

The main result of our analysis is that it gives us some information on
topology change at the quantum level. This is true for geons as well as
for particles on the plane \cite{syr}. Our algebra ${\cal A}^{(1)}$
has to do with large distance observations. In other words,
we can only probe low energy aspects of the theory. We will argue in
Section \ref{S5} that geons, i.e, handles in the plane,  can 
be created and annihilated  in a quantum fashion as a
consequence of the scale of observations. We would like to mention
that other types of topology change, like creation of baby 
universes, do not fit naturally in our framework and will not be
considered here. 

One advantage of the algebraic approach is that we can do this
analysis without going into the details of the ``complete'' underlying
field theory. We can determine the  spectrum ${\hat {\cal A}^{(1)}}$ of the 
geons, i.e., the set of possible irreducible representations of ${\cal
  A}^{(1)}$, but a particular field theory may restrict the available
possibilities in ${\hat {\cal A}^{(1)}}$. The determination of these
possibilities 
requires the study of  particular examples of the underlying 
field theories. That may be a very difficult task. In this paper our intention
is to use the simplified algebraic ``field'' theory and see what it can teach
us. It is remarkable that such a simple framework can reveal important
features of quantum geons such as a constraint involving  spin and statistics 
as well as rules for quantum topology change. The former connection is
investigated in another paper \cite{us}.

An approach similar to ours is explored in reference \cite{KML}. Its
author views the geon as a vortex-antivortex pair, in which case the
algebra describing it {\em is} a
quantum double. This description does not consider the internal
diffeomorphisms of the geon, as it aims to describe vortices on a
two-dimensional surface with handles. Accordingly, in \cite{KML}, the
setting is a two-dimensional surface $\Sigma _{g,n}$ of {\it genus}
$g$ {\em and $n$ punctures}, whereas in this work we consider a $2$-surface
$\Sigma _{g,0}$ of {\it genus}
$g$ {\em without punctures}. Our approach is also
different inasmuch as we are interested in considering ``large
diffeormorphisms'', i.e., elements of the mapping class group of
$\Sigma _{g,0}$. More
specifically, in \cite{KML}, the topology of this surface is a passive
background where a theory of pointlike vortices is
defined, and its author only deals with diffeomorphisms moving particles
(punctures) around or through handles. To us, the geons (handles)
themselves, {\em including their internal structure}, are
the entities of interest. The diffeomorphisms moving these handles
are the ``large diffeomorphisms'' we mentioned above. As an
illustration of the above mentioned differences, in
a typical process considered in \cite{KML}, one can make a test vortex go
through or around a handle, whereas in our case one can conceive of
``test geons'' going through other handles. Our procedure allows a
natural generalization towards quantum gravity, which is the
issue of another paper \cite{us}. 

We recall the notion of topological geons in Section \ref{S2}. A
special emphasis is given to orientable geons in $(2+1)d$. The field algebra
is described in Section \ref{S3}. 
Section \ref{GASP} gives the effective description of a geon as seen from a 
large distance. The relevant subalgebra ${\cal D}\subset {\cal A}^{(1)}$  
is the same as for a point particle. The representations of 
${\cal D}$ will play an important role when we discuss topology.
Quantization of
the system is given in Section \ref{S4}. In this section we are able
to classify the irreducible representations for a class of algebras
${\tilde {\cal A}}$ that includes our algebra of interest as a
particular example. It is worthwhile to point out that the field algebras
for vortices in $(2+1)d$ and for rings in $(3+1)d$
considered in \cite{syr} are also examples of ${\tilde {\cal
    A}}$. Section \ref{S5} describes how topology can change in this
quantum theory, as a consequence of the scale of observation. We end
with some concluding remarks and prospects of future work.

\sxn{Topological Geons}\label{S2}

The term {\em geon} was used for the first time by J.A. Wheeler
\cite{Wheel} to designate a lump of electromagnetic energy held
together by its own gravitational field, forming a spatial region of
non-zero curvature, typically very small. In the  context of this
paper, however, this term will have a wholly different meaning, namely
it will signify a {\em topological} geon, a soliton-like
excitation in topology first discussed by Friedman and Sorkin
\cite{FS}, and whose properties were further  elaborated by many
authors \cite{FW,Aneziris,SS,Sam}. In this section,
we review the definition and basic properties of topological geons,
and refer the reader to the literature for further details.  

We start with some basic preliminary definitions. Let 
$M_{1}$,$M_{2}$ be connected $n$-dimensional topological
manifolds, possibly with boundaries. We define their connected sum, 
$M_{1} \# M_{2}$, as follows: take $n$-balls $B^n_{i}$ in the
interiors of $M_{i}$ ($i=1,2$) and remove their interiors. We thereby
add spheres ${\bf S}^{n-1}$ to the boundaries of $M_{1}$ and 
$M_{2}$. Now identify the points of these spheres via a
homeomorphism. The resulting manifold is 
$M_{1} \# M_{2}$. If $M_{1}$ and $M_{2}$ are oriented,
we further require that this homeomorphism be orientation reversing so
that $M_{1} \# M_{2}$ is also oriented. 
It follows trivially  from the definition that $M\# {\bf S^n}$ is
homeomorphic to
$M$ itself. The connected sum of two tori is shown on Fig. 2.1.

\begin{figure}
\centerline{\epsfbox{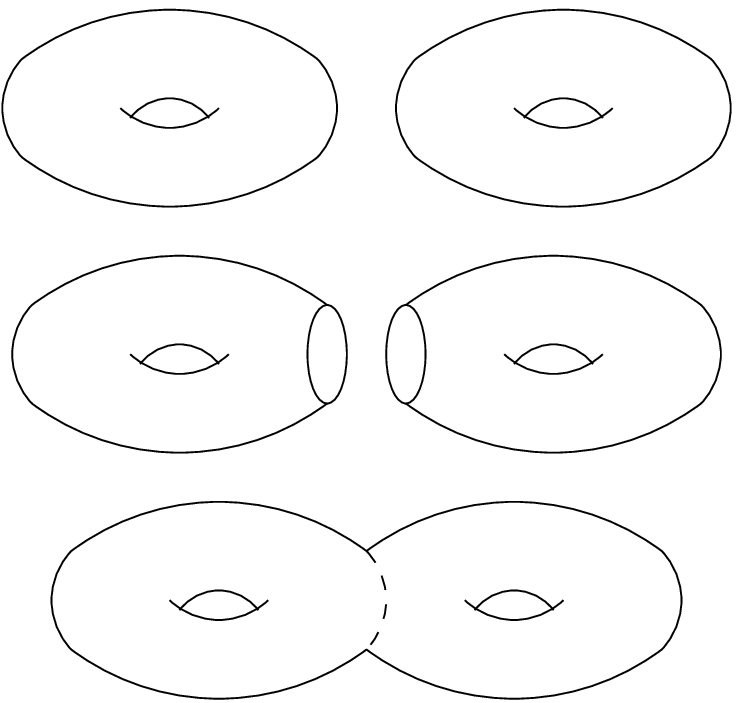}}
{\small {\bf Fig. 2.1:}
The connected sum of two tori ${\bf T}^{2}$. One first removes
a disc from each torus and then glues them along the new boundaries.}
\end{figure} 

In this paper we shall be interested in a decomposition of spacetime
by spacelike hypersurfaces (spatial manifolds). In dealing with
gravity, one is usually interested in spacetime metrics which induce
an asymptotically flat (or asymptotically conical, in the $(2 + 1)d$
case)Riemannian metric on each hypersurface. For a 
certain ``frozen time'' $t$, the hypersurface $S_{t}$ should therefore be
topologically a manifold with one {\em asymptotic region}, i.e., there
exists a compact region $R_{t}\subset S_{t}$ whose complement in
$S_{t}$ is homeomorphic to ${\IR}^{n}\backslash B^{n}$, where $(n+1)$ is
the dimension of spacetime and $B^{n}$ is the standard $n$-ball in
${\IR}^{n}$. In Fig. 2.2 one can see $(2 + 1)d$ oriented geons  (which
are nothing but handles on a plane, see below) on a $2$-dimensional spatial
slice.   This motivates the following definition: an $n$-manifold is
said to have one asymptotic region iff it is homeomorphic to
$R^{n} \# M$, where $M$ is a closed (i.e., compact and boundaryless),
connected $n$-manifold. Typical cases
of interest are 2 and 3 manifolds with one asymptotic region , which
are to be thought as spatial slices of $(2+1)$- and $(3+1)$-dimensional
spacetimes respectively.

 \begin{figure}
\begin{center}
\begin{picture}(0,0)
\put(250,15){$S$}
\end{picture}
\epsfbox{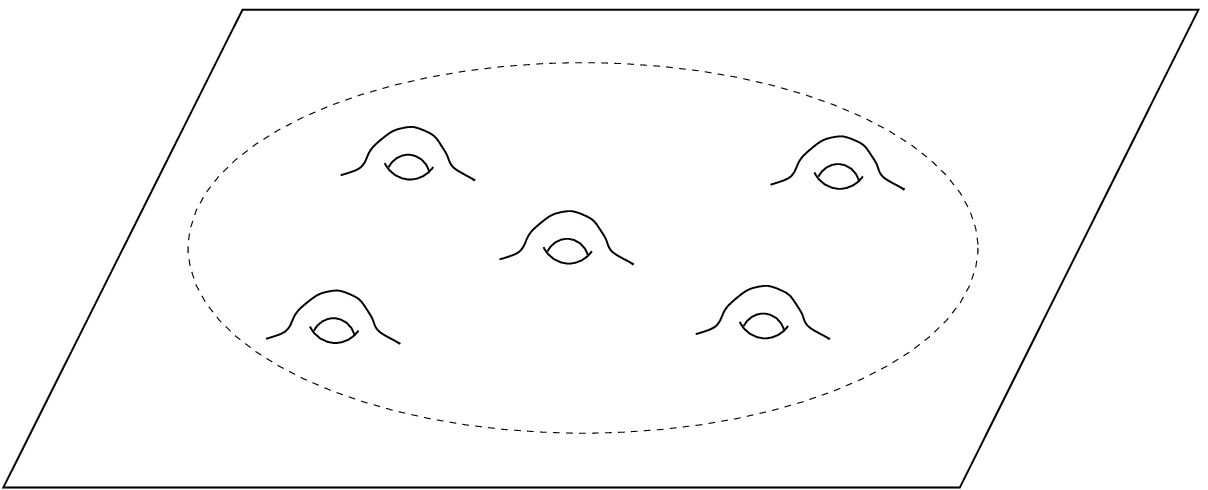}
\end{center}
{\small {\bf Fig. 2.2:}
A plane with a finite number of geons (handles) is an example of
a manifold with one asymptotic region. Note that
all topological complexity can be localized within a circumference
$S$, and the geons can be isolated from each other. Outside $S$, one
has simply a flat plane.}
\end{figure}

In 2 and 3 dimensions it is known \cite{Still,Ham} that there exists a class
of closed connected manifolds ${\cal P}_{i}$ called {\em prime}
manifolds. An $n$-manifold $M$ is said to be prime iff $M = 
M_{1} \# M_{2}$ implies that one of $M_{1}$, $M_{2}$ is
  an $n$-sphere. One can prove that given any compact $n$-manifold ($n=2,3$) 
$M$, there exists a unique decomposition
\be
  M = {\cal P}_{1} \# ... \# {\cal P}_{N},
\ee
where ${\cal P}_{i} \neq {\bf S^{n}}$. Uniqueness means (apart from
some technicalities - see ref. \cite{Ham}) that
given another decomposition ${\cal P}'_{1} \# ... \# {\cal P}'_{N'}$, we
have $N=N'$ and (after possible reordering) ${\cal P}_{i}$ is homeomorphic to
${\cal P}'_{i}$. Each prime component of $M$ is called a topological geon.

In 2 dimensions, ignoring ${\bf S}^{2}$ from consideration,the only
 prime manifolds are ${\bf T}^{2}$,  which
is orientable,  and the ``cross cap'' ${\bf RP}^{2}$, which is non-orientable 
 \cite{Still}. In this paper we will consider only orientable geons, therefore 
we will have to deal only with ${\bf T^{2}}$.  
Connected sums with ${\bf S}^{2}$ are clearly immaterial. In 3 dimensions
there are  infinitely many
prime manifolds, only partially classified. As examples we can give 
the 3-torus ${\bf T^{3}}$ and the ``handle'' ${\bf S^{2}} \times {\bf S^{1}}$.

From the aforementioned prime decomposition it is clear that any $n$-manifold
$M$ \mbox{($n=2,3$)} with an asymptotic region can be decomposed
as 
\be
M={\IR ^{n}} \# {\cal P}_{1} \# ... \# {\cal P}_{N}.
\ee
Now consider
${\IR}^{n} \# {\cal P}_{i}$. One can always find an $n-1$ sphere in
${\IR^{n}} \# {\cal P}_{i}$ whose interior contains ${\cal P}_{i}$. By a
suitable choice of the metric this region can be thought of to be as small as
one pleases, i.e., the topological complexity can be localized (for
details see ref. \cite{Aneziris}). In 2 spatial dimensions this means
that one is allowed to put the handle inside of a circle and suppose
the radius of the circle to be very small. Then one has a very small handle
surrounded by a vast flat plane. It is in this sense that we refer to
the geon as ``soliton like'' at the beginning of this section: just as
a soliton corresponds to a localized excitation of some field, outside
of which one has the vacuum, the geon
is a localized excitation of the topology itself, the ``vacuum'' in
this case being the flat space (see Fig. 2.3). In general, since  
${\cal P}_{i}$ is
prime,  one may say
that it represents an elementary topological excitation. We therefore
say that ${\IR}^{n} \# {\cal P}_{i}$ is a space with one geon. The
manifold 
${\IR}^{n} \# {\cal P}_{1} \# ... \# {\cal P}_{N}$ is therefore
seen as  a space
 with $N$ geons. These prime manifolds attached to ${\IR ^{n}}$ can be
 isolated from one another in the same way as we localized one
 single geon \cite{Aneziris}, and for many purposes one can think of
 geons as particles. Again, in 2 spatial dimensions one  can have many
isolated small handles.
\begin{figure}
\begin{center}
\begin{picture}(0,0)
\put(250,65){$S$}
\end{picture}
\epsfbox{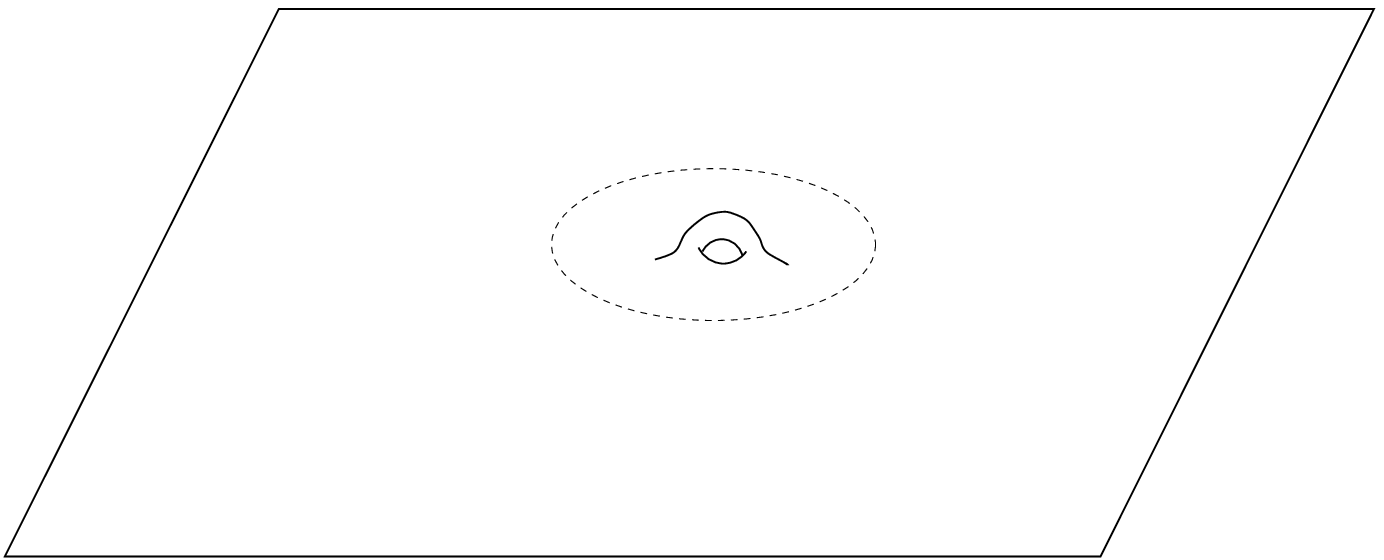}
\end{center}
{\small {\bf Fig. 2.3:} A ``solitary'' geon can be seen as a localized excitation,
or ``soliton'', of topology. From a distance it looks like a point
particle.}
\end{figure} 

The importance of geons to us lies in the fact that, as long as we
preserve connectivity and consider a space manifold with one
asymptotic region, {\em topology change amounts to creation and
  annihilation of geons}. Henceforth we restrain our attention to the
case when the space is 2 dimensional, connected and oriented, with one
asymptotic region. We assume, furthermore, that connectivity and 
orientability are preserved during topology change. Although somewhat
restrictive, this case is still of much interest. Our assumptions
imply, on the other hand, that the geons of interest will be those
associated to copies of ${\bf T}^{2}$, i.e., topology changing processes
will mean creation and annihilation of handles on a plane. As we will
see below, creation and annihilation will have for us a meaning different from
the usual geometrical one. Instead they will be related to what a
``distant'' observer will be able to measure from a quantum
theoretical standpoint.

\sxn{The Field Algebra for $(2+1)d$ Topological Geons }\label{S3}

Our aim in this section is to define some ``observables'' which
describe the topological character of a geon. However, the term
``algebra of observables'' to designate the algebra describing geons
would actually be a misnomer, for as we will see shortly, this algebra
includes operators which cannot be observables. To describe geons, we
will use the low-energy limit of a field theory in their presence. In
this limit, the theory becomes topological, and therefore provides us with
quantities capable of probing the topological features of the
background, and hence the geons. The kind of algebra
which we will encounter is composed by a part related to the fields,
via their holonomies around non-contractible paths, and to physical
operations (some of them not observable locally) which may change
these holonomies. This algebra is what is known in the literature as a
{\em field algebra} (for a detailed definition, see for instance \cite{field}).

We will follow an approach  inspired by the work of the 
Amsterdam group, which is reported in ref. \cite{prop}. In this work, 
the group investigates the properties of topological solutions of a 
$(2+1)d$ gauge field theory in Minkowski spacetime
where the gauge symmetry of a Lie group $G$ is spontaneously 
broken to a finite group $H$ by a non-vanishing expectation value of a
Higgs field $\Phi$. We shall briefly review their discussion,
referring the reader to \cite{prop} for details. The Lagrangian is given by 
\be
{\cal L} \, =\, \frac{1}{4} F^{a}_{\mu\nu} F^{\mu\nu}_{a} \, +\, 
Tr[(D_{\mu} \Phi)^{*} \cdot (D^{\mu} \Phi )] \, -\, V(\Phi ) \; ,\label{higgs} 
\ee
where $\mu ,\nu =0,1,2$, and $a$ is a Lie algebra index. For
simplicity, we assume that $G$ is connected and simply connected.
The fields $F^{a}_{\mu\nu}$ are the 
components of the field strength of the Yang-Mills potential 
$A_{\mu}^{a}$ and $D_{\mu}$ denotes the covariant derivative
determined by this potential. The Higgs field $\Phi$ is in the adjoint
representation and can be expanded in
terms of generators $T^a$ of the Lie algebra of $G$, and $V(\Phi )$ is
a $G$-invariant potential.
In this paper we shall be concerned with the low energy, or
equivalently, the long range behavior of this theory, in the
temporal gauge $A_{0}^{a} = 0$. This is obtained
by  minimizing the three
terms in the energy density separately. Minimizing the term
corresponding to the energy density of the Yang-Mills field, we obtain
the condition $F^{a}_{\mu\nu} = 0$, from which we conclude that we
are dealing only with flat
connections. The minimum of the potential restricts the values of the
Higgs field to the vacuum manifold, which is invariant
by $H$. Finally, the condition $D\Phi =0$, required for minimizing the
energy density from the second term, tells us that the holonomies
\be
\tau (\gamma )= 
P\exp \{ \int_{\gamma} A^{a}_{i} T_{a} ds^i \} \; ; \; i \in \{1,2\} \; ,
\label{flat}
\ee
take values in the finite group $H$. 

Here and in what follows we will
fix a base point $P$ for loops, so that all loops will begin and end
at $P$.    

This gauge theory may have topologically non-trivial, static solutions
such as vortices. It is very well known that the core radii of these
vortices are inversely proportional to the mass of the Higgs boson,
and therefore they may be viewed as point-like in the low-energy
regime of the theory. Hence, according to a standard argument, to
describe the $N$-vortex solutions we may consider solutions for
the vortex equations 
\bea
\label{static}
F^{a}_{ij} &=& 0 ; \nonumber \\
D_i\Phi &=& 0 ; \nonumber \\
V(\Phi) &=& 0,
\eea
on a spacetime of the form $\Sigma \times \IR$, where $\Sigma$ is the
plane with $N$ punctures, playing the role of the vortices. Now, take
a solution $(A,\Phi)$ for the vortex equations (\ref{static}). By fixing a
point $P \in \Sigma$, the holonomy of $A$ around any closed path $\gamma$
based at $P$ depends only on its homotopy class, since $A$ is flat.
It takes values into a subgroup $H$ of $G$, which preserves the vacuum
manifold, in view of the equations for $\Phi$ \cite{prop}. Therefore,
any solution of the vortex equations determines a
homomorphism $\tau$,
\be
\label{homo}
\tau :\pi _1(\Sigma ) \rt H,
\ee
between the fundamental group $\pi_1 (\Sigma )$ and the group
$H$. Conversely, given such a homomorphism $\tau$ we can define a
solution for eqs.(\ref{static}) in the following way. Take the
universal covering space $\tilde{\Sigma}$ of $\Sigma$. It is the total
space of a principal bundle over $\Sigma$ with structure group $\pi_1
(\Sigma)$. Via the homomorphism $\tau$ we can construct an associated
principal $H$-bundle over $\Sigma$, which is a subbundle of the
original $G$-bundle. Since $H$ is finite, this bundle
has a unique flat connection $A^{a}_{i}$, which can be viewed as a reducible
connection on the $G$-bundle. We now find a $\Phi$. By
fixing some $\Phi _0$ in the vacuum manifold, we have that, since
$\Phi$ must be covariantly constant, we define $\Phi(P) = \Phi_0$ and
its value can be obtained for
each $x \in \Sigma$ by parallel transporting $\Phi _0$ along some path
from $P$ to $x$ in $\Sigma$: 
\be
\label{transport}
\Phi(x)= 
P\exp \{ \int_{P}^{x} A^{a}_{i} T_{a} ds^i \} \Phi _0. 
\ee
The pair ($A^{a}_{i}$, $\Phi$) thus constructed is obviously a
solution of the vortex equations. Therefore the space of solutions for
the vortex eqs. (\ref{static})
is essentially parametrized by homomorphisms $\tau : \pi_1(\Sigma)
\rightarrow H$. Each such homomorphism is then a vortex configuration.

Let us take the example of a single point-like vortex on the
plane. The non-contractible loop $\gamma $ that encircles the
singularity has holonomy  
$\tau (\gamma )$ equal to the flux carried by the vortex. 
In trying 
to capture only topological information, one is not concerned with the 
position of the vortex, but only with its flux, characterized by some 
group element $h\in H$. In other words, the ``configuration space'' $T$ 
for the vortex is labeled by elements of $H$. 
Hence, the algebra ${\cal F}(H)$ 
of complex-valued functions on $H$ with pointwise multiplication plays
the role of ``position observables''. Let us denote by $P_h\in {\cal
F}(H)$, the characteristic function supported at $h\in H$. Then
\be
\label{dP}
P_h~P_{h'}=\delta _{hh'}P_h.
\ee

In terms of homomorphisms, we have not yet exploited all the degrees
of freedom the system has. Indeed, we have that $H$ can act by conjugating all holonomies based at $P$ by an element of
$H$. Hence, a flux
$\sigma$ transforms under $H$ as
\be
\sigma\, \mapsto \, h \sigma h^{-1} \; ,\label{gauge}
\ee
In other
words, we have an action of $H$ by conjugation of the fluxes. We shall
simply refer to this action as the $H$-transformations. The group
elements $h \in H$ can be regarded as operators, also denoted
by $h$, acting 
on the functions  $f\in {\cal F}(H)$ via (\ref{gauge}). In other
words,
\be
\label{dg}
hP_{\sigma}h^{-1}=P_{h \sigma h^{-1}}.
\ee
The multiplication of two $H$-transformations is the same as the
group multiplication. Therefore the algebra of such 
operators turns out to be the group algebra ${\IC }(H)$.

As for the physical interpretation of the
$H$-tranformations we note that the mathematical action depicted in
(\ref{gauge}) is entirely equivalent, from a physical standpoint, to
what occurs when one makes a vortex of flux $\sigma$ encircle a source
of flux $h$ at infinity. Since such operation is non-local, one must
conclude that the $H$-transformations cannot be
considered local in the theory, i.e., cannot be implemented by local
operators.  
    
The field algebra is then the semi-direct product 
$D(H)={\IC}(H)\semidirect {\cal F}(H) $, the so-called Drin'feld 
double. It has the  structure of a
quasi-triangular Hopf algebra. The Hopf structure \cite{kassel} 
means in particular the existence of a co-product, i.e, a map 
\[
\Delta \, :\, D(H)\,  \longrightarrow \,  D(H)\otimes D(H) \; ,
\]
which is a homomorphism of algebras (and with further properties to be
discussed in Section 4). In \cite{prop} the fluxes are
seen as particles in $(2 + 1)d$ and are then first quantized: the (internal)
Hilbert space ${\cal H}$ is constructed, and the
elements of the algebra $D(H)$ act as operators on this Hilbert space.
${\cal H}$ decomposes into irreducible representations of $D(H)$,
corresponding to the different particle sectors of the quantum
theory. The existence of a co-product allows one to understand fusing processes between
particles. The quasi-triangularity implies the existence of the
$R$-matrix, $R \in D(H)\otimes D(H)$, responsible for all braiding 
processes between particles. For further details see \cite{prop}. 

How is the topology of $\Sigma$ taken into account in this
approach? First of all, we have seen that the physically distinct
vortex configurations are in one-to-one correspondence to the space of
conjugacy classes of homomorphisms of $\pi_1(\Sigma)$ into
$H$. Moreover, it is well known that for a finite group $H$ the latter
space is in one-to-one correspondence with equivalence classes of
principal $H$-bundles over $\Sigma$ \cite{Steenrod}. Therefore the only degree of
freedom in this theory is the topology of these bundles
\cite{Dij,freed}. Second, a configuration for which the holonomy is
trivial around some puncture is indistinguishable, from the standpoint
of the low-energy theory, to another in which that particular puncture
is absent. Therefore the low-energy theory somehow actually allows for
``topology fluctuations'' of $\Sigma$ as long as we stay within its
limits, and as far as ``creation and annihilation'' of punctures is
concerned.     

In order to determine the field algebra for  a topological 
geon, we will try to follow a method similar to the one used for
vortices in the plane, 
respecting carefully the differences between the two systems. 
We will first try to find the analogues of the ``position 
observables'' for a geon. Now, $\Sigma$ is the plane with one or more
handles, and for simplicity we shall assume throughout the that there
are no vortices, i.e., we work in the zero vortex number sector of the
low-energy limit of the theory given by the Lagrangian in
(\ref{higgs}). This is in constrast with \cite{KML}, where vortices
are the central interest. There, the vortices determine the state of a
handle, whereas in the present work all non-trivial configurations will be
related solely to holonomies around and through the handles. In other
words, the geons are our main concern, and the background field theory
merely defines their states.

\begin{figure}[t]
\begin{center}
\begin{picture}(0,0)
\put(90,-10){$P$}
\put(93,2){$\ast$}
\put(50,50){$\gamma_1$}
\put(100,50){$\gamma_2$}
\put(190,50){$\gamma_3$}
\end{picture}
\epsfbox{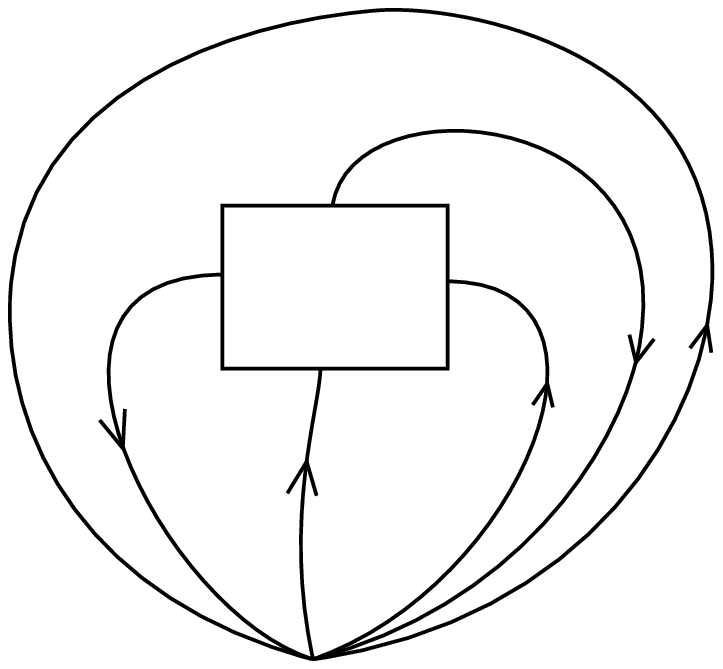}
\end{center}
{\small {\bf Fig. 3.1:} The figure shows the loops $\gamma_i$ ($1 \leq i \leq
  3$). The homotopy classes $[\gamma_1]$ and $[\gamma_2]$ generate the
  fundamental group. The class $[\gamma _3]$ is not independent of
  $[\gamma_1]$ and $[\gamma_2]$.}
\end{figure} 

Let us start by taking $\Sigma $ to be the plane with a handle. On all
figures, a geon will be thought of as a square hole on the plane,
with the opposite sides identified. One can show that 
$\pi _1(\Sigma )$ has two generators $[\gamma _1]$ and $[\gamma _2]$, shown by 
Fig. 3.1. It can be shown that 
\[
[\gamma _3]=[\gamma_1][\gamma _2][\gamma_1]^{-1}[\gamma _2]^{-1}.
\]
Actually, $\pi _1(\Sigma )$ is freely generated by
$[\gamma _1]$ and $[\gamma _2]$. Let 
$g=W([\gamma _1],[\gamma _2])\in \pi _1(\Sigma ) $, be a word in 
$[\gamma _1],[\gamma _2]$ and their inverses.
Then $\tau $ maps $g$ to  
$W(a,b)\in H $ where 
$a=\tau (\gamma _1)$ and $b=\tau (\gamma _2)$.  Therefore the map 
$\tau :\pi _1(\Sigma )\rt H$ is completely 
characterized by the fluxes $\tau (\gamma _1)=a$ and 
$\tau (\gamma _2)=b$. Since there is no relation between $a$ and $b$, the set 
$T$ of all maps is labeled by $H\times H $. 

{\bf Definition}:  Let $H$ be a finite group and $\Sigma$ the plane with one 
geon, i.e., a two dimensional  manifold given by 
\[
M\, =\, {\bf R}^2 \, \# \, {\bf T}^2 \; .
\]
Let $\gamma _1$ and $\gamma _2$ denote representative loops whose classes 
generate $\pi_1 (\Sigma )$. 
We define a  classical configuration $\tau _{(a,b)}\in T$ 
of a geon as the homomorphism defined by
\be
\tau _{(a,b)} (\gamma _1)=a,\mbox{ and } \tau _{(a,b)}(\gamma _2)=b.
\ee

It is important to bear in mind that $T \cong H\times H$ and therefore 
that it is a finite discrete set. 
For simplicity of notation, a geon configuration will be denoted 
simply by a pair $(a,b)$ of fluxes. Note that we are not explicitly
identifing those configurations which differ by an
$H$-transformation. This is because wave functions need only be
``covariant'' under the symmetries of the problem, and only its
modulus squared and other observable quantities, like Aharonov-Bohm
phases, must be invariant. In our approach, this will happen
naturally, just as in \cite{prop}.

With $T\cong H\times H$ being the configuration space for a geon, the 
corresponding algebra of ``position observables'' is ${\cal
  F}(T)$. Instead of working with the abstract algebra, we specify a
representation. Let $V$ be the (finite-dimensional) complex vector
space generated by the vectors    
$\mid a,b\rangle , a,b \in H$. We will call the representation on
$V$, to be defined below, the defining representation.
The algebra ${\cal F}(T)$ 
is generated by projectors on $V$ denoted by $Q_{(a,b)}$. They
are defined by  
\be
\label{projectors}
Q_{(a,b)} \mid c,d \rangle \, =\, \delta_{a,c} \, \delta_{b,d} 
\mid c,d \rangle \; .
\ee
The operator $Q_{(a,b)}$ represents a ``delta 
function'' supported at $(a,b)$, i.e., it gives $1$ when evaluated on $(a,b)$,
and zero everywhere else. Indeed, from (\ref{projectors}) one finds that
\be
\label{algebraF}
Q_{(a,b)} Q_{(c,d)} \, =\,  \delta_{a,c} \delta_{b,d}  Q_{(c,d)} \; ,
\ee

Besides the projectors $Q_{(a,b)}$, which play the role of
position operators in ordinary quantum mechanics, we have also some
operators capable of changing $(a,b)$. They are somewhat 
analogous to momentum operators. For example, like in the case of vortices,  
$H$-transformations act on the configurations. 
It turns out that for a geon there are additional operators besides 
$H$-transformations. They correspond to the 
action of the group  $Diff ^\infty(\Sigma ) $ 
of diffeomorphisms of $\Sigma $ that keeps infinity invariant. 

We will start by first examining the $H$-transformations. 

The group $H$ acts on $T$ simply by conjugating both fluxes in $(a,b)$. This 
will induce an operator $\hat \delta _g$ for each $g\in H$, acting on the defining 
representation $V$ by 
\be
\hat \delta _g \mid a,b \rangle \, =\, \mid gag^{-1} ,gbg^{-1} \rangle. 
\label{del}
\ee
From (\ref{del}) one sees that the multiplication of operators 
$\hat \delta _g$ is given by 
\be
\label{groupalgebra}
\hat \delta_{g} \hat \delta_{h} \, =\, \hat \delta_{gh} \; .
\ee
The corresponding algebra generated by $\delta _g$ is the group algebra 
${\IC}(H)$. The relation between ${\cal F}(H\times H)$ and ${\IC}(H)$ 
can be derived from (\ref{projectors}) and (\ref{del}). One sees immediately
that 
\be
\label{actionofH}
\hat \delta _g Q_{(a,b)}\hat\delta ^{-1} _g \, =
\, Q_{(gag^{-1},\, gbg^{-1} )} \; .
\ee
In other words, the algebra ${\IC}(H)$ acts on ${\cal F}(H\times H)$.

Besides $H$-transformations, fluxes $(a,b)$ can change under the action of 
the group $Diff ^{\infty}(\Sigma )$. 
It is clear that elements belonging to the subgroup 
$Diff ^{\infty}_0 (\Sigma )$, the component connected to identity,
 act trivially on $\pi _1(\Sigma )$ \footnote{For simplicity, we take
$P$ to be at infinity. Even if we do not, the holonomies will be
invariant under the action of $Diff ^{\infty}_0 (\Sigma )$.} and hence on 
$(a,b)$. Therefore what matters is the action of the so-called mapping
class group $M_\Sigma $ \cite{birman,balachandran}, defined as
\be
\label{mcg}
M_{\Sigma } \, =\, 
\frac{Diff^{\infty} (\Sigma )}{Diff^{\infty}_{0}(\Sigma )} \; .
\ee
\begin{figure}[t]
\centerline{\epsfbox{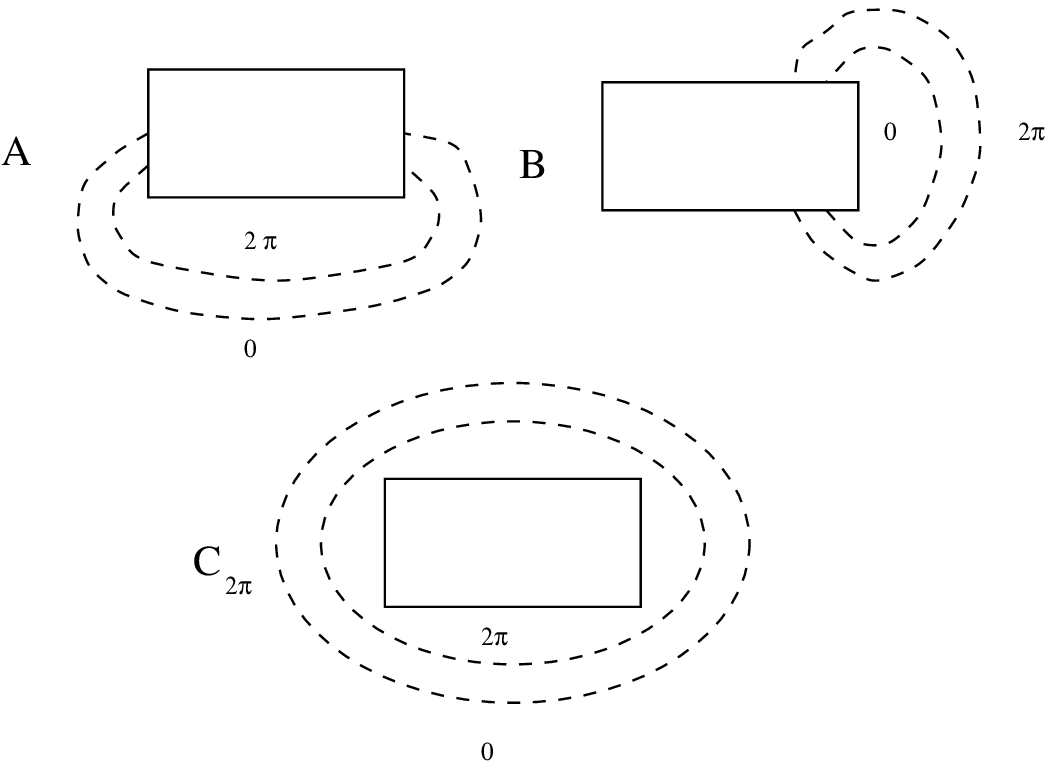}}
{\small {\bf Fig. 3.2:} Dehn twists corresponding to diffeomorphisms of the mapping
  class group. The annuli enclose loops, which we have
  omitted in the figure. Rotations are counterclockwise by convention.}
\end{figure}

For the present case, $\Sigma $ is the plane with a single geon and the mapping
class group is isomorphic to the central extension of the group
$SL(2,{\bf Z})$, denoted by $St(2,{\bf Z})$ and called the Steinberg
group.  This is the same as the mapping class
group of a torus minus one point \cite{Aneziris}.
We denote generators of $M_{\Sigma }=St(2,{\bf Z}) $ by $A$ and $B$. They 
correspond to (isotopy classes of) diffeomorphisms \footnote{One can
  see from (3.14) that the mapping class group consists of isotopy
  classes of diffeomorphisms. Throughout this paper we shall loosely
  use a representative in a class as the class itself.} called Dehn
twists. 
A Dehn twist is realized as follows. Take a loop in $\Sigma$. Then
draw an annulus enclosing the loop and 
introduce radial coordinates $r\in [0,1]$, with $r=0$ and $r=1$ 
corresponding to the boundaries of the annulus, see Fig. 3.2. 
Then rotate the points of the 
annulus in such a way that the angle of rotation $\theta (r)$ is zero
for $r=0$ 
and gradually increases, becoming $2\pi $ at $r=1$. 
Figure 3.2 shows how to produce Dehn twists, and in Fig. 3.3, we show
how the
Dehn twist $B$ deforms the loop $\gamma_1$.
There is also the Dehn twist along a loop enclosing the geon, which
can be interpreted as the $2 \pi$-rotation of the geon
\cite{erice,FS,Aneziris}. This Dehn twist will be important when we
discuss the notion of spin of a topological geon.
The corresponding annulus is denoted by $C_{2\pi}$ in
Fig. 3.2. However, $C_{2\pi}$ is not independent of $A$ and $B$.  
One can show that \cite{Aneziris}
\be
C_{2\pi}=(AB^{-1}A)^{4}. 
\ee
The group $M_\Sigma $ is generated by $A$ and $B$, 
with the relation that $C_{2\pi}$ commutes with $A$ and $B$. It is
useful to think of
the elements of $M_\Sigma $ as words $W(A,B)$ in $A$, $B$ and their inverses. 
 
The action of $A$ and $B$ on $[\gamma _i]\in \pi_{1} (\Sigma )$ 
induces an action on $(a,b)\in T$, and therefore induces operators 
$\hat A$ and $\hat B$ in the defining representation acting on
$V$. Let us take as an 
example the action of $B$ on $\gamma _1$, as given by Fig. 3.3. One sees that 
$[\gamma _1] \rt [\gamma _1][\gamma _2]$, and therefore $a\rt ab $. On the 
other hand, $B$ keeps $[\gamma _2]$ invariant.
\begin{figure}[t]
\begin{center}
\begin{picture}(0,0)
\put(80,-7){$P$}
\put(83,3){$\ast$}
\end{picture}
\epsfbox{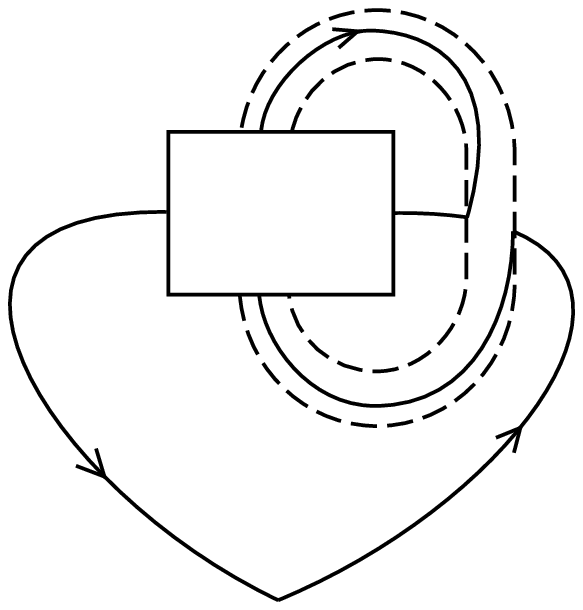}
\end{center}
{\small {\bf Fig. 3.3:} Dehn twist $B$ and its action on $\gamma_1$.}
\end{figure}
One can verify that $A$ and $B$ induce the following operators:
\bea
\label{abc}
\hat A \mid a,b\rangle \, & =&\, \mid a, ba \rangle \; ,\nonumber\\ 
\hat B \mid a,b\rangle \, &=&\, \mid ab,b\rangle \; .
\eea
For an arbitrary word $W(A,B)$, the corresponding operator is 
$W(\hat A,\hat B)$, i.e., the
same word but with $A$ and $B$ replaced by $\hat A$ and $\hat B$. 
For example, the Dehn  twist $C_{2\pi}$ of Fig. 3.2 
is written as $(AB^{-1}A)^4$ and the
corresponding operator $\hat C_{2\pi}$ can be immediately computed to be
\be
\label{opC}
\hat C_{2\pi} \mid a,b\rangle \, =\, 
\mid {\bf c}^{-1}a{\bf c} ,{\bf c}^{-1}b{\bf c}\rangle \;,
\ee
where ${\bf c}=aba^{-1}b^{-1}$. 

It is also possible to perform rotations 
of the  geon by integer multiples of the angle $\frac{\pi}{2}$ 
using $C_{\frac{\pi}{2}} =AB^{-1}A$. The corresponding operator is given by
\be
\label{piover2}
\hat C_{\frac{\pi}{2}} \mid a,b\rangle \, =\, 
\mid b^{-1},bab^{-1}\rangle \, . 
\ee

The group ${\cal M} $ generated by $\hat A$ and $\hat B$ defined by 
(\ref{abc}) is the one relevant for defining the field algebra.   
Contrary to the mapping class group, ${\cal M}$ is a finite group. It turns 
out that an infinite number of words $W(\hat A,\hat B)$ is equal to the 
identity operator and that ${\cal M}$ can be naturally identified with 
$M_{\Sigma }$ divided by a certain normal subgroup. 

Let $M_0$ be a subgroup of $St(2,{\bf Z})$ defined
as
\[
M_0 \, =\, \left\{ h \in St(2,{\bf Z}) \mid \hat h \mid a,b \rangle =
\mid a,b \rangle \; , \; \forall a,b \in H \right\} \; .
\]
It is easy to see that $M_0$ is a normal subgroup. In fact, given any
word $W\in St(2,{\bf Z})$, such that 
\[
W(\hat A,\hat B)\mid a,b\rangle \, =\, \mid a' ,b' \rangle \; ,
\]
we have the relation
\bea
W^{-1}(\hat A,\hat B)\, \hat h\, W(\hat A,\hat B) 
\mid a,b\rangle \, &=&\, W^{-1}(\hat A,\hat B)\, \hat h \mid a' ,b' \rangle \, 
=\, W^{-1}(\hat A,\hat B) \mid a' ,b' \rangle \, =\nonumber \\
&=& \mid a,b\rangle \; .\nonumber
\eea
We define the {\em effective} mapping class group ${\cal M}$ acting on the
defining representation $V$ as the quotient
\[
{\cal M} \, =\, St (2,{\bf Z}) /M_0 \; .
\]

We now show that  ${\cal M}$ is finite.  Let $n$ be the order of $H$ and 
$a_i,\,\,i=1\ldots n$ its elements. We construct a basis for  $V$ as 
\[
{\cal B}=\, \left\{ \mid a_i ,a_j \rangle \; , \;\,\,\, 
i,j =1\ldots n \right\} \; .
\]

The group ${\cal M}$ acts as a subgroup of the permutation 
group of  the elements in ${\cal B}$, thus the order of ${\cal M}$ is at 
most equal to $n^2!$.

The algebra generated by the operators $\hat A$ and $\hat B$ is the group 
algebra ${\IC}({\cal M})$.  Together with ${\IC}(H)$ and 
${\cal F}(H\times H)$ it gives us the total field algebra ${\cal A}^{(1)}$
for a  single topological geon. 
From the definitions (\ref{projectors}), 
(\ref{del}) and ({\ref{abc}}) one sees that 
\[
\hat\delta _g\hat A=\hat A\hat\delta _g, \,\,\,\,\,\hat\delta _g\hat
B=\hat B\hat\delta _g,
\]
\[
\hat\delta _gQ_{(a,b)}\hat\delta _g^{-1}=Q_{(gag^{-1},gbg^{-1})},
\]
\[
\hat C_{2\pi} \hat A = \hat A \hat C_{2\pi}, \,\,\,\,\, \hat C_{2\pi}
\hat B = \hat B \hat C_{2\pi},
\]
\be
\hat AQ_{(a,b)}\hat A^{-1}=Q_{(a,ba)},\,\,
\hat BQ_{(a,b)}\hat B^{-1}=Q_{(ab,b)}.\label{relations} 
\ee
Therefore, both algebras ${\IC}(H)$ and ${\IC}({\cal M})$ act on 
${\cal F}(H\times H)$.  The action of a generic word
$W(\hat A,\hat B)$ on $Q_{(a,b)}$ will be denoted by
\be
\label{Waction}
W(\hat A,\hat B)Q_{(a,b)}W^{-1}(\hat A,\hat B)= Q_{(w^{(a)},w^{(b)})}.
\ee
where $(w^{(a)},w^{(b)})$ is a pair of words in $a$ and $b$ and their 
inverses, representing the action of $W(A,B)$ on $(a,b)$. 

There are two equivalent ways of presenting ${\cal A}^{(1)}$. One is by using 
the defining representation of (\ref{projectors}), (\ref{del}) and ({\ref{abc}}).
Another way is to define  ${\cal A}^{(1)}$ as the algebra generated by 
$Q_{(a,b)}$, $\hat\delta _g$, $\hat A$ and $\hat B$ with the relations
(\ref{relations}). In any case, we have that 
\be
\label{productinA}
{\cal A}^{(1)} = \IC(H\times{\cal M}) \semidirect {\cal F}(H\times H).
\ee

We shall now introduce the field algebra 
for two topological geons following exactly the same ideas as for a single 
topological geon. We recall that for a single geon, ${\cal A}^{(1)}$ 
consists of three sub-algebras, namely  the ``position
observables''  ${\cal F}(T)$, the $H$-transformations 
${\IC}(H)$,  and the ``translations'' , i.e., a realization 
${\cal M}$ of the mapping class group $M_\Sigma $. The algebra 
${\cal A}^{(2)}$ for two geons 
will consist of the same three distinct parts, with 
$T=H\times H\times H \times H \equiv H^4$ and $\Sigma $ replaced by a
plane with two handles.

We shall  start by examining the fundamental group 
\[
\pi _1(\Sigma )=\pi_1 ({\bf R}^2\# {\bf T}^2 \# {\bf T}^2).
\]
Let $\gamma _i,\, i=1,2,3,4$ be the loops shown by Fig. 3.4.
One can show that $\pi _1(\Sigma )$ is the  free group generated by 
$[\gamma _i]$. A ``configuration'' $\tau $ of two topological geons is 
given by a homomorphism $\tau :\pi _1(\Sigma )\rt H $. Therefore $\tau $ is 
completely characterized by the holonomies $\tau (\gamma _i)\in H$ along the 
loops $\gamma _i$. Since there are no relations among $[\gamma _i]$'s, 
the holonomies $\tau (\gamma _1),\tau (\gamma _2),\tau (\gamma _3)$ and 
$\tau (\gamma _4)$ are four arbitrary elements of $H$. 
In other words, the set $T^{(2)}$ of configurations $\tau $ can be 
identified with $T^{(1)}\times T^{(1)}=(H\times H)\times (H\times H)$, where 
$T^{(1)}$ is the configuration space for a single geon. The corresponding 
algebra ${\cal F}(H^4)$ 
is thus the direct product of the algebra of single geons,i.e.,
\[
{\cal F}(H^4)\cong {\cal F}(H\times H)\otimes {\cal F}(H\times H).
\] 
\begin{figure}
\centerline{\epsfbox{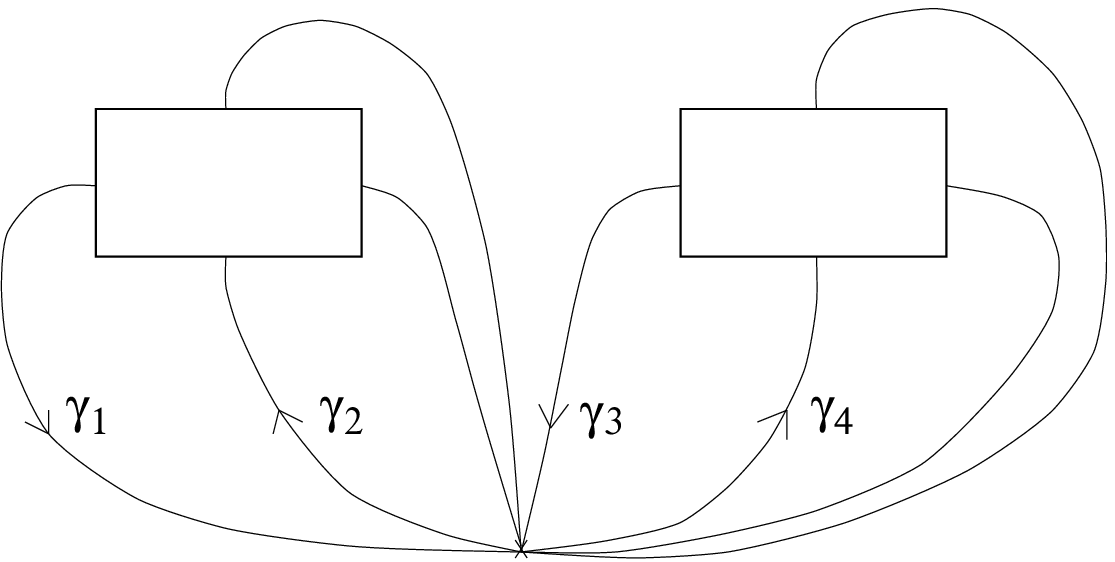}}
{\small {\bf Fig. 3.4:} The homotopy classes  $[\gamma_i ]$ ($1\leq i\leq 4$)
relative to the loops shown in the figure generate the fundamental
group of plane with two geons.}
\end{figure}

It is natural to work with the defining representation on $V\otimes V$ 
spanned by vectors of the form
\[
\mid a_1 , b_1 \rangle \otimes \mid a_2 , b_2 \rangle \; ,
\]
where the subscripts denote the respective geons. The ``position
observables'' are generated by projectors $Q_{(a_1 ,b_1 )} \otimes
Q_{(a_2 ,b_2 )}$ acting on $V\otimes V$ in the obvious way, i.e., 
\be
Q_{(a_1 ,b_1 )} \otimes Q_{(a_2 ,b_2 )}
\mid a'_1 , b'_1 \rangle \otimes \mid a'_2 , b'_2 \rangle =
\delta_{a_1,a'_1}\delta_{b_1,b'_1}\delta_{a_2,a'_2}\delta_{b_2,b'_2}
\mid a_1 , b_1 \rangle \otimes \mid a_2 , b_2 \rangle \,.
\label{position}
\ee
Therefore, the  ``position''  operators belong to 
${\cal A}^{(1)}\otimes {\cal A}^{(1)}$.

The action of $H$-transformation $g\in H$ on the fluxes 
$(a_1,b_1,a_2,b_2)$ is by a global conjugation. This induces the action
\be
\mid a_1 , b_1 \rangle \otimes \mid a_2 , b_2 \rangle \, \rt \
\mid ga_1g^{-1} , gb_1g^{-1} \rangle \otimes 
\mid ga_2g^{-1} , gb_2g^{-1} \rangle  
\ee
on $V\otimes V$. The corresponding operator is obviously identified with  
$\hat \delta _g\otimes \hat \delta _g \in  {\IC}(H)\otimes {\IC}(H)$, since 
\be
\hat \delta _g\otimes \hat \delta _g
\mid a_1 , b_1 \rangle \otimes \mid a_2 , b_2 \rangle=
\mid ga_1g^{-1} , gb_1g^{-1} \rangle \otimes 
\mid ga_2g^{-1} , gb_2g^{-1} \rangle .
\label{2gauge}
\ee
Hence, $H$-transformation operators  also belong to 
${\cal A}^{(1)}\otimes {\cal A}^{(1)}$.

We now start to consider the action of the mapping class group $M_\Sigma $.
For two or more geons, $M_\Sigma $ is much more 
complicated than for a single geon \cite{birman}. The mapping class group
is generated by Dehn twists of the type $A$ and $B$ (see Fig. 3.2) for each 
individual geon together with diffeomorphisms involving pairs of geons.  

Let $A_i,B_i$, $i=1,2$ be the generators of the ``internal diffeos'' for each 
individual geon. The corresponding operators acting on $V\otimes V$ 
are clearly given by
\[
\hat A_1=\hat A\otimes \II, \,\,\,\,\, \hat A_2=\II \otimes \hat A 
\]
\be
\hat B_1=\hat B\otimes \II, \,\,\,\,\, \hat B_2=\II \otimes \hat B
\label{internal}
\ee
where $\II $ is the identity operator on $V$. 

There are two additional classes of 
transformations besides the internal diffeos. The first one, called 
exchange, is the analogue of the elementary braiding of two 
particles. The second, called handle slide, has no analogue for particles,
since it makes use of the internal structure of the geon.

So far, all operators in the algebra for ${\cal A}^{(2)}$ were of the form 
$x\otimes y\in {\cal A}^{(1)}\otimes {\cal A}^{(1)}$. It turns out that 
this is not the case for exchanges and handle slides.
They correspond somewhat to 
interactions and cannot be written strictly in terms of operators in
${\cal A}^{(1)}\otimes {\cal A}^{(1)}$. In order to describe interactions between geons, 
we need to define a pair of flip automorphisms of $V \otimes V$. They
are necessary
in the construction of the exchange and handle slide operators. 

{\bf Definition}: Given a two geon state 
\[
\mid a_1 , b_1 \rangle \otimes \mid a_2 , b_2 \rangle \in V\otimes V\; ,
\]
the flip automorphisms $\sigma $ and $\gamma $ are defined by:
\bea
\sigma \mid a_1 , b_1 \rangle \otimes \mid a_2 , b_2 \rangle \, &:=&\, 
\mid a_2 , b_2 \rangle \otimes \mid a_1 , b_1 \rangle \; , \nonumber\\
\gamma \mid a_1 , b_1 \rangle \otimes \mid a_2 , b_2 \rangle \, &:=&\, 
\mid a_1 , b_2 \rangle \otimes \mid a_2 , b_1 \rangle \; . \nonumber
\eea

Both are not given geometrically as morphisms of the mapping class
group, but unless one introduces these operators, the algebra of two
geons cannot be related directly to the algebras for a single
geon. We will show that the algebra ${\cal A}^{(2)}$ can be obtained from
the tensor product ${\cal A}^{(1)}\otimes {\cal A}^{(1)}$ when we add
$\sigma $ and $\gamma $.

In the exchange process, two geons permute their positions. 
In our convention, the geon
on the right (left) moves counterclockwise to the position of the
left(right) (see
Fig. 3.5).
\begin{figure}
\centerline{\epsfbox{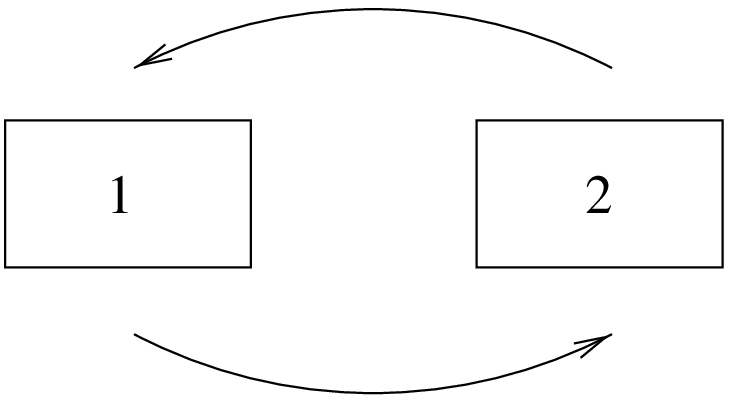}}
\begin{center}
{\small {\bf Fig. 3.5} Geon exchange.}
\end{center}
\end{figure}
The effect of a geon exchange on the states is of the form
\be
\label{geonexchange}
{\cal R} \mid a_1 , b_1 \rangle \otimes \mid a_2 , b_2 \rangle \,
=\, \mid {\bf c}_{1}^{-1} a_2 {\bf c}_{1},
{\bf c}_{1}^{-1} b_2 {\bf c}_{1}\rangle \otimes 
\mid a_1 , b_1 \rangle \; ,
\ee
where ${\bf c}_{1} = a_1b_1a^{-1}_1b^{-1}_1$. This operator is
equivalent to braiding operators for particles and 
also satisfy the Yang-Baxter equation,
\be
\label{yb}
({\cal R} \otimes {\II })({\II } \otimes {\cal R})
({\cal R} \otimes {\II }) \, =\, ({\II } \otimes {\cal R})
({\cal R} \otimes {\II })(\II \otimes {\cal R}) \; .
\ee
One can verify that the exchange operator (\ref{geonexchange}) 
may be written as the product
\be
\label{R}
{\cal R}=\sigma \, R
\ee
where $R\in {\cal A}^{(1)}\otimes {\cal A}^{(1)}$ is the analogue of the 
universal $R$-matrix for a quasi-triangular Hopf algebra. In our case $R$ is
given  by
\be
\label{R'}
R=\sum _{a,b}Q_{(a,b)}\otimes {\hat \delta}^{-1} _{aba^{-1}b^{-1}} \, . 
\ee

The handle slide is shown in Fig. 3.6. 
In (a), the geon is viewed as a rectangular box on the plane. 
In (b), we have identified two edges of the rectangle and the
geon is represented as two circles on the plane connected by dotted lines. 
The handle slide is defined as the
operation that performs a double counterclockwise exchange of the 2nd
and 3rd circles followed by a clockwise $2\pi$-rotation of each one of them. 
\begin{figure}
\begin{center}
\epsfbox{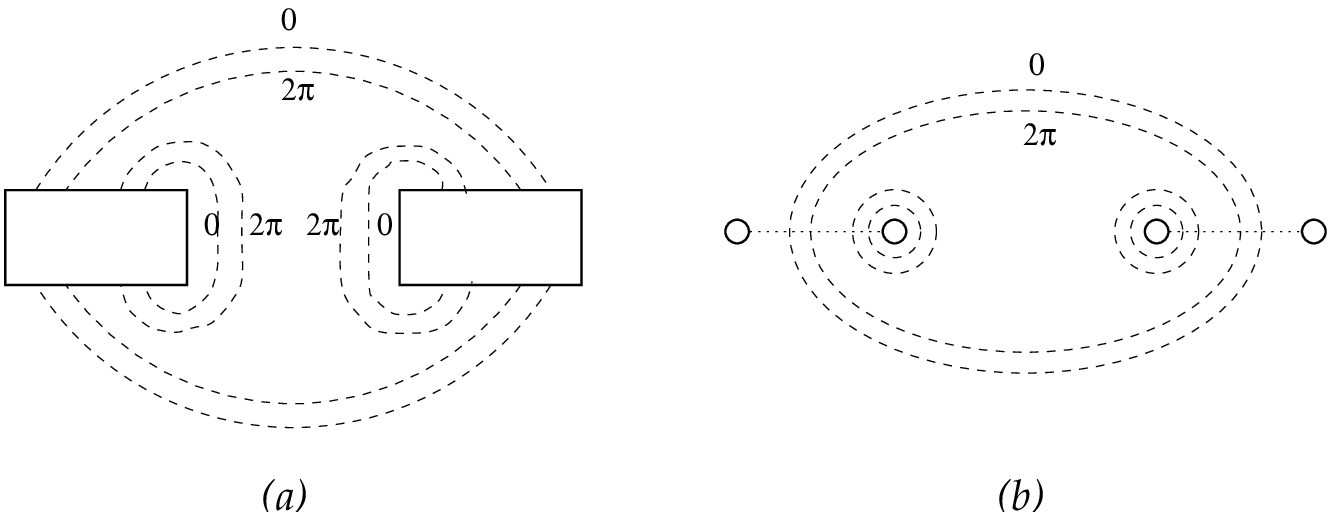}
\end{center}
{\small {\bf Fig. 3.6:} The handle slide is interpreted geometrically as
  the full monodromy of two handles followed by a rotation of
  $2\pi$ of each handle. The figure shows two equivalent
  representations for the handle slide: In (a), the geon is
  viewed as a rectangular box on the plane. In (b), we have identified
  two edges of the rectangle and the geon is represented as two circles
  on the plane. } 
\end{figure}
\begin{figure}
\begin{center}
\begin{picture}(0,0)
\put(65,-7){$(c)$}
\put(240,-7){$(d)$}
\put(65,117){$(a)$}
\put(240,117){$(b)$}
\end{picture}
\epsfbox{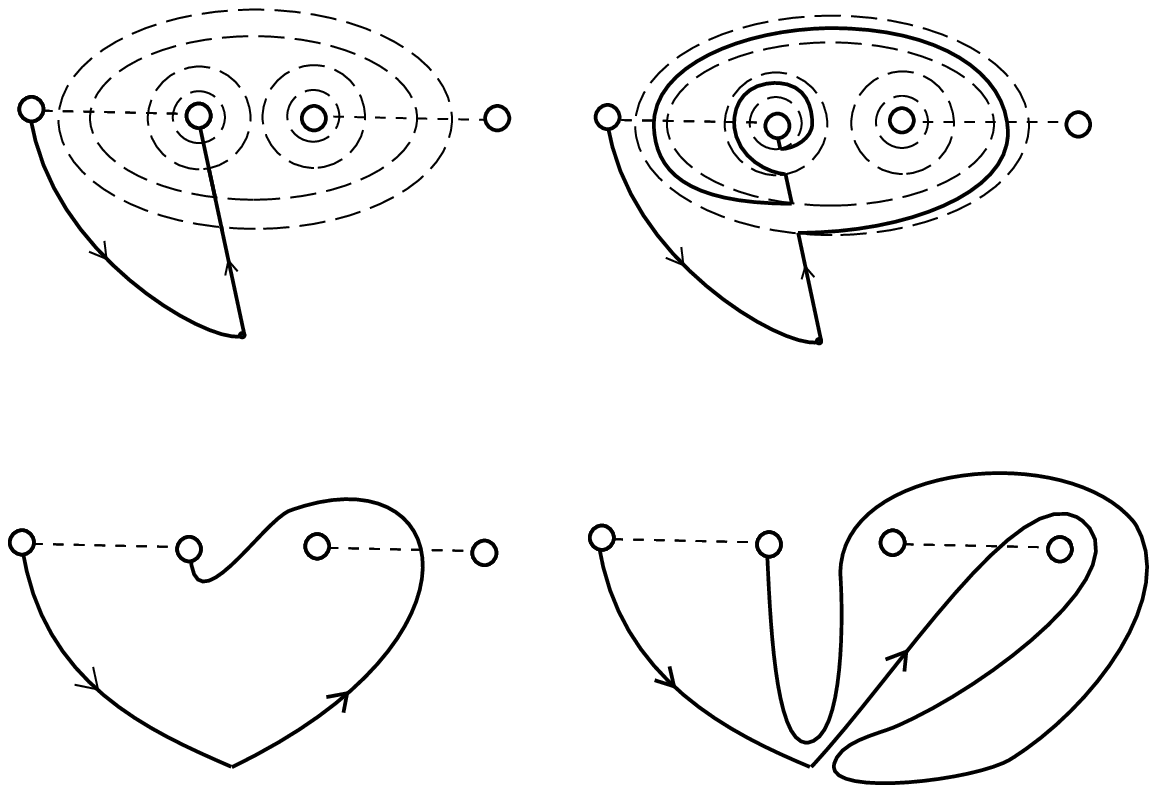}
\end{center}
{\small {\bf Fig. 3.7:} Figure (a) shows the loop $\gamma _1$ defined on
  Fig. 3.4. The transformed loop $\tilde \gamma _1$ is indicated in
  (b). Figures (c) and (d) are two steps in the deformation of
  $\tilde \gamma _1$.} 
\end{figure}
As expected, this Dehn twist acts on the generators $[\gamma _i]$ of 
$\pi _1(\Sigma )$ given in Fig. 3.4, and therefore on the holonomies.  
Under the action of the transformation indicated in Fig. 3.6, the 
loops $[\gamma _i]$ will be mapped into new loops 
$[\tilde \gamma _i]$. As an example let us show how the handle slide
acts on $\gamma _1$. 
The loop $\gamma _1$ is shown in
Fig. 3.7 (a). After the action of the diffeo, $\gamma _1$ is 
mapped to $\tilde \gamma _1$, indicated in Fig. 3.7 (b). We need to
express $\tilde \gamma _1$  in terms of the generators $[\gamma _1]$. 
It is easy to see that $\tilde \gamma _1$ can be deformed to 
$\gamma _1 \alpha \gamma _4$, where $\alpha $ is the loop enclosing
the second geon. The sequence of deformations is indicated by
Fig. 3.7 (b), (c) and (d). As $\alpha $ measures the total flux 
$a_2b_2a_2^{-1}b_2^{-1}$, it is easy to see that $\tilde \gamma _1$
will measure the flux $a_1 (a_2 b_2 a_{2}^{-1})$. One can repeat the
same procedure for the other loops and show that the action on the 
loops induces an action on $V\otimes V$ given by the
following handle slide operator ${\cal S}$:
\bea
\label{handleslide}
& & {\cal S}  \mid a_1 , b_1 \rangle \otimes \mid a_2 , b_2 \rangle = 
\nonumber \\ 
& &= 
\mid a_1 (a_2 b_2 a_{2}^{-1}) , 
(a _2 b_2 a_{2}^{-1})^{-1} b_1 (a_2 b_2 a_{2}^{-1})
\rangle \, \otimes
\mid (a_2 b_2 a_{2}^{-1})^{-1} b_1 (a_2 b_2 a_{2}^{-1}) a_2 , b_2 
\rangle \; .
\eea

This is a very complicated action on states, but there is a way to
write ${\cal S}$ as a product of elements of 
${\cal A}^{(1)} \otimes {\cal A}^{(1)}$ with flip
automorphisms in the same way as the operator ${\cal R}$. The  result is
\bea
\label{H}
{\cal S}\, &=&\, \gamma \left( \II  \otimes
\sum_{g,h}Q_{(g,h)} \hat\delta _g \right) \left( \II  \otimes B \right)
\gamma \cdot \nonumber \\
&\cdot & \left( \II  \otimes C_{\frac{\pi}{2}} \right) \gamma  
\left( \II  \otimes
\sum_{g,h}Q_{(h,g)} \hat\delta _{g^{-1}} \right) \gamma \cdot \nonumber\\
&\cdot & \gamma \left( \II  \otimes C_{\frac{\pi}{2}} \right) 
\gamma \left( B^{-1} \otimes \II  \right) \gamma 
\left( \II  \otimes C_{-\pi} \right). \; 
\eea

This completes the description of ${\cal A}^{(2)}$. The algebra for two geons
is generated by the elements of ${\cal A}^{(1)}\otimes {\cal A}^{(1)}$, 
${\cal R} $ and ${\cal  S}$.

These constructions can be easily generalized to write down the algebra
${\cal A}^{(n)}$ for $n$ geons. It is clear that 
\[
\underbrace{ {\cal A}^{(1)} \otimes \cdots \otimes {\cal A}^{(1)} }_{n} 
\, \subset \,{\cal A}^{(n)}  \; .
\]
The complete algebra ${\cal A}^{(n)}$ can be obtained by adding 
the operators ${\cal R}_{ij}$ and ${\cal S}_{ij}$ of 
exchange and handle slide between
the $i$-th and the $j$-th geons. They can be easily constructed by using 
operators analogous to (\ref{R}) and (\ref{H}), acting 
on the $i$-th and $j$-th entries of $V\otimes ... \otimes V$. 

It is clear that the elements $A_i$, $B_i$, ${\cal R}_{ij}$ and ${\cal
  S}_{ij}$ of ${\cal A}^{(n)}$ generate, 
under multiplication, a group ${\cal M}_n$ that is 
homomorphic to the mapping class group $M_\Sigma $ for $n$ geons.
Besides the relations proper to $M_\Sigma $, however, we will have
  extra relations so that ${\cal M}_n$ becomes effectively  finite.

\sxn{The Geon as a Single Particle}\label{GASP}

We have seen up to now that a geon is a topological object with
internal structure. In quantum theory, it can be described 
by the algebra ${\cal A}^{(1)}$. However,
for a large distance observation, we may 
disregard the operators that probe its internal structure and
describe it by a subalgebra ${\cal A}^{(1)}_L$.
In this 
approximation, a topological geon seems to be  no different from a particle on 
the plane, or a vortex in $(2+1)d$. We may guess that the large
distance field algebra ${\cal A}^{(1)}_L$ is an algebra equivalent to
$D(H)$, the quantum double introduced in Section \ref{S3}. Actually
this is not exactly true. We will see that ${\cal A}^{(1)}_L$ for a
single geon has extra elements besides the ones corresponding to
$D(H)$.

Long distance observables should not see the internal structure of the
geon. For instance, in performing Aharonov-Bohm-type experiments in
this long-distance scale, one should expect to see only the
effects of the total flux, or the holonomy of the large loop 
${\bf \gamma _3}$. Therefore, the only detectable projector in this
scale is the one with support at the total flux ${c}$ of a single
geon. It is naturally defined as
\be
\label{qc}
Q^{(1)}_{c} \, :=\, \sum_{a,b} \delta_{aba^{-1}b^{-1} ,{c}} 
Q_{(a,b)} \; .
\ee
The index $(1)$ in $Q^{(1)}_{c}$ is to remind us that this large
distance projector is an element of 
${\cal  A}^{(1)}$, the algebra of a single geon.

The algebra of operators $Q^{(1)}_{c}$ can easily be obtained from the
algebra (\ref{algebraF}), resulting in
\be
\label{subalgebra}
Q^{(1)}_{{c}_1} Q^{(1)}_{{c}_2} \, =\, \delta_{{c}_1 , {c}_2 }
Q^{(1)}_{{c}_1} \; .
\ee
Hence, the algebra generated by $Q^{(1)}_c$ is isomorphic to ${\cal F}(H)$.

The $H$-transformation operators $\hat\delta _g\in {\IC}(H)$ act on
$Q^{(1)}_{c}\in  {\cal F}(H)$. From (\ref{relations}) and (\ref{qc}), 
one can verify that
\be
\label{subgauge}
\hat\delta _gQ^{(1)}_{c}{\hat\delta _g}^{-1}=Q^{(1)}_{g{c}g^{-1}}.
\ee
Therefore $\hat\delta _g$ has to be regarded as a large distance
operation. To make the notation uniform, we define
\be
\hat\delta ^{(1)}_g:=\hat\delta _g \, .
\ee

The operators $Q^{(1)}_c$  
should commute with local operators in ${\cal A}^{(1)}$, namely the diffeos
$\hat A$ and $\hat B$. This  must be true since the action of the mapping
class group cannot change $[\gamma _3]$. That is because one can make 
$\gamma _3$ very large, such that the Dehn twists $A$ and $B$ do not act on
$\gamma _3$. See Fig. 3.1 and Fig. 3.2.  In fact, one can verify that
\be
\hat AQ^{(1)}_c\hat A^{-1}=\,\,\,\hat BQ^{(1)}_c\hat B^{-1}=Q^{(1)}_c.
\ee

Let us call ${\cal D}^{(1)}\subset {\cal A}^{(1)}$ the algebra 
generated  by $Q^{(1)}_c$ and ${\hat\delta}^{(1)}_g$. 
It is clear that ${\cal D}^{(1)}$ is 
isomorphic to the Drin'feld quantum double 
$D(H)\cong {\cal F}(H)\otimes {\IC }(H)$. As a consequence, ${\cal
  D}^{(1)}$ has the structure of a 
quasi-triangular Hopf algebra \cite{kassel}. In this paper we will be
interested mostly in two properties of a quasi-triangular Hopf
algebra, namely the existence of a co-product and the universal $R$ matrix. 

A co-product on ${\cal D}^{(1)}$ is a linear map 
\[
\Delta : {\cal D}^{(1)} \, \longrightarrow \, {\cal D}^{(1)} 
\otimes {\cal D}^{(1)} \; ,
\]
which is co-associative,
\[
(\Delta \otimes Id ) \circ \Delta \, =\, (Id \otimes \Delta )\circ
\Delta \; ,
\]
and a morphism of algebras,
\[
\Delta (a\cdot b) \, =\, \Delta (a) \cdot \Delta (b) \; .
\]
For the quantum double, the co-product has the expressions
\be
\label{coproduct1}
\Delta (Q^{(1)}_c) \, =\, \sum_{g} 
Q^{(1)}_{g}\otimes Q^{(1)}_{g^{-1}c}  \; .
\ee
and
\be
\label{coproduct2}
\Delta ({\hat\delta }^{(1)}_g)={\hat\delta }^{(1)}_g\otimes {\hat\delta} ^{(1)}_g.
\ee

The quasi-triangularity of the quantum double implies the existence of an
$R$-matrix, which is responsible for the exchange processes. The $R$-matrix
for the quantum double can be written as
\be
\label{Rsub}
{\cal R}^{(1)} \, =\sigma \, \sum_{g\in H} Q^{(1)}_g  \otimes 
{\hat\delta}^{(1)}_{g^{-1}} \; .
\ee

We recall that the full algebra ${\cal A}^{(1)}$ also has an $R$-matrix given
by (\ref{R}). One should ask whether the $R$-matrix (\ref{Rsub}) 
for the subalgebra ${\cal D}^{(1)}\subset {\cal A}^{(1)}$ is compatible with 
(\ref{R}). It is a simple matter to show that they are actually identical.

We may think of the $R$-matrix for  ${\cal A}^{(1)}$ as a trivial 
extension of the $R$-matrix of ${\cal D}^{(1)}$. An important question is 
whether it is also possible to extend the co-product to the 
entire algebra  ${\cal A}^{(1)}$. We have 
reasons to believe that $\Delta $ cannot be extended. One reason is that
the co-product is related to fusion of particles at the quantum level,
which is physically reasonable. However, it is harder to imagine that
two handles put together could be seen as a single handle. 

Another large distance element in $A^{(1)}_{L}$ is the operator 
$C^{(1)}$ responsible for the
Dehn twist on a cycle that encloses the entire geon. In other words,
$C^{(1)}$ is the $2\pi$-rotation of the geon: 
\be
C^{(1)} \equiv C_{2\pi }.
\ee
Note that $C^{(1)}$ commutes with all elements of ${\cal
  D}^{(1)}$. Since $C_{2\pi }^N=\II$ for some $N$, it generates a group algebra
isomorphic to $\IC(\IZ_N)$. 

Summarizing, the long distance algebra ${\cal A}_L^{(1)}$ is isomorphic  to 
$D(H)\otimes \IC(\IZ_N)$. In other words, on a large distance scale, a
geon is equivalent to a particle with a frame.

Consider next the two-geon configuration and its corresponding algebra ${\cal
  A}^{(2)}$. The associated long distance algebra can be visualized as
  follows.  
Let the two geons shrink to localized objects and at the same
time approach each other. 
At the end a point-like object will remain
and we should look for the operators that still make sense in the
limit. It is clear that such operators  will
be a) the total flux projector $Q^{(2)}_c$ of the two geons; b) the
  $H$-transformations and c) the Dehn twist around a cycle enclosing both geons. 

The projection operator for the total flux of the system is given by
\be
Q^{(2)}_{c} \, :=\, \sum_{a,b,a',b'}
\delta_{aba^{-1}b^{-1}a'b'{a'}^{-1}{b'}^{-1} ,{c}} Q_{(a,b)}\otimes
Q_{(a',b')} \; .
\ee
The index $(2)$ indicates that $Q^{(2)}_c$ is an element of
${\cal A}^{(2)}$.
One can write this expression in a more transparent way as follows:
\be
\label{q2}
Q^{(2)}_c=\sum _gQ^{(1)}_g\otimes Q^{(1)}_{g^{-1}c}.
\ee
Similarly, the $H$-transformation is given by
\be
\label{d2}
\hat\delta _g ^{(2)}:=\hat\delta ^{(1)}_g\otimes \hat\delta ^{(1)}_g.
\ee
If we compare the last two equations with the definition 
(\ref{coproduct1})-(\ref{coproduct2}), we see that
\be
\label{Q2}
Q^{(2)}_c=\Delta (Q^{(1)}_c),
\ee
\be
\label{g2}
{\hat\delta }^{(2)}_g=\Delta (\hat\delta ^{(1)}_g).
\ee

Let us denote by ${\cal D}^{(2)}$ 
the algebra generated by $Q^{(2)}_c$ and $\hat\delta ^{(2)}_g$. From (\ref{Q2})
and (\ref{g2}) it follows that ${\cal D}^{(2)}$ is homomorphic to 
${\cal D}^{(1)}$. Actually, it is a simple matter verify that they are 
isomorphic. 

As in the previous case, the long distance algebra ${\cal A}^{(2)}_L$ has an
extra generator given by the Dehn twist $C^{(2)}$ on a cycle enclosing
both geons, with
\be
C^{(2)}={\cal R}^2.
\ee
As one would expect, the algebra ${\cal A}^{(2)}_L$ is isomorphic to 
$D(H)\otimes \IC(\IZ_N)$ and therefore also describes a
particle with a frame.

It is clear now what is the long distance algebra ${\cal A}^{(n)}_L$
for $n$ geons. It is generated by the Dehn twist $C^{(n)}$ on a cycle
enclosing the $n$ geons, together with elements $Q_c^{(n)}, \hat\delta ^{(n)}_g
\in {\cal A}^{(1)}\otimes ...\otimes {\cal A}^{(1)}$ given by the
iterative application of the co-product. For example, for $n=3$,
\bea
Q_c^{(3)} & = &(Id\otimes \Delta)\otimes \Delta (Q^{(1)}_c) ,\\
\hat\delta _g^{(3)} &=& (Id\otimes \Delta)\otimes \Delta (\hat\delta ^{(1)}_g).
\eea   
Notice that because of the co-associativity property, we could have written 
$(\Delta \otimes Id)\otimes \Delta$ instead of $(Id\otimes \Delta)\otimes \Delta$
in the last two formulae.

\sxn{Quantization}\label{S4}

The algebra ${\cal A}^{(1)}$ describes the topological degrees of freedom for
a single geon on the plane. To quantize the system we need to find an 
irreducible representation of ${\cal A}^{(1)}$ on a Hilbert space
${\cal H}$. However, this Hilbert space will branch into irreducible
representations of the field algebra:
\be 
{\cal H} = \oplus_{r}{\cal H}_{r},
\ee
where ${\cal H}_{r}$ denotes a particular irreducible representation
describing a certain geon type.
The algebra is finite dimensional, and therefore there will be a
finite number of  
irreducible representations of ${\cal A}^{(1)}$. Furthermore, the Hilbert 
spaces ${\cal H}$ are all finite dimensional. Each representation gives us a
possible one-geon sector of the theory.

In the case of quantum doubles, the irreducible representations
are fully classified. See for 
instance ref. \cite{SV}. For the case of geons, the algebra is more
complicated because of the existence of internal
structure. Nevertheless, the representations of ${\cal A}^{(1)}$ are quite 
similar to the ones of the quantum double of a finite group. This is
not totally surprising,
since in a certain limit, as discussed in the previous
section, we recover the quantum double ${\cal D}^{(1)}\cong D(H)$. Actually,
we can define a class of  algebras ${\cal A}$ that can have its representations
classified and that are generic enough to contain the quantum double 
and the algebra ${\cal A}^{(1)}$ as particular cases.
In the spirit of \cite{SV}, one can then get all representations  of
${\cal A}$. 

\noindent
{\bf Definition}: Let $X$ be a finite set and $G$ a finite group acting on 
$X$. In other words, there is a map $\alpha _g:X\rt X$ for each $g\in G$.  
As usual, we denote by ${\cal F}(X)$ the algebra of functions on $X$ and 
by $\IC (G)$ the group algebra of $G$.  
We define the algebra ${\cal A}$ as the vector space 
\[
{\cal A}:={\cal F}(X)\otimes \IC (G)
\]
with basis elements denoted by $(Q_{x},g)$, $Q_{x} \in {\cal F}(X)$ and $g
\in \IC (G)$, and the multiplication 
\be
\label{prodinA}
(Q_x , g ) \cdot (Q_y ,h )\,:=\, 
(Q_x Q_{\alpha_{g} (y)} , g h ) \; .
\ee
Here, $Q_x$ is the characteristic function supported at $x \in X$. Let
$x_0$ be an element of $X$. We denote by  $K_{x_0}\subset G$ the 
stability subgroup with respect to $x_0$, i.e.,
\be
K_{x_0} \, =\, \left\{ g\in G \mid  \alpha_g (x_0 )=x_0 \right\} \; .
\ee

The stability subgroup $K_{x_0}$ divides  the group $G$
into equivalence classes of left cosets. Let $N$ be the number of
equivalence classes and let us choose 
a representative $\xi_i \in G$, $i=1\ldots N$ for each class,  with the 
convention that  $\xi_1 =e$. We can write the following partition of
$G$ into left cosets: 
\be
G\, =\, \xi_1 K_{x_0} \cup \xi_2 K_{x_0} \cup \ldots \cup \xi_N K_{x_0} \; .
\ee

Let us point out that $\IC(G)$ seen as a vector space carries a left
representation of $G$, the action of $G$ being by left product. All irreducible
representations of $G$ can be obtained by reducing this
representation. In particular, any vector space carrying an
irreducuble representation (IRR) of $K_{x_0}$ can be viewed as a
subspace of $\IC(G)$. 

Note that ${\cal F}(X)$ plays a dual role: it is an algebra, but it is
itself a vector space which is acted upon 
by the group $G$, according to $gQ_x := Q_{\alpha_{g} (x)}$. This
can be extended to an action of $\IC(G)$ in the obvious way. Also, it
acts upon itself by left (pointwise) product. In what
follows we shall denote the elements $Q_x$ by $\mid x \rangle$
whenever we want to view it as a vector belonging to the
representation of $\IC(G)$ on ${\cal F}(X)$ just defined. In this
``passive'' role it is acted upon, instead of acting on some
representation of the algebra of functions. 

We can now state the following result.   

\noindent
{\bf Theorem }
Let $\mid j \rangle_{\rho}$, $j=1\ldots n$ be a basis of a subspace $V_\rho$ of
$\IC(G)$ carrying an IRR $\rho$ of $K_{x_0}$. Then, for (a fixed) $x_0 \in X$, 
elements $\xi_i \in G$, $i=1\ldots N$ and $\mid j \rangle_{\rho} \in
{\IC} (G)$, $j=1\ldots n$ as stated above, the vectors
\[
\xi_i \mid x_0 \rangle \otimes  \mid j \rangle_{\rho} :=
\mid \alpha_{\xi_i}(x_0) \rangle \otimes  \mid j \rangle_{\rho} \; ,
\]
form a basis for an IRR of the algebra 
${\cal A}$, given by 
\[
(Q_x , g )\mid \alpha_{\xi_i}(x_0) \rangle \otimes  \mid j
\rangle_{\rho} := \delta_{x, \alpha _{\xi_{i'}}(x_0)}\mid \alpha_{\xi_{i'}}(x_0)
\rangle \otimes\Gamma^{(\rho)}(\beta)_{kj} \mid k \rangle_{\rho}\; ,
\]
where $\xi_{i'}$ and $\beta$ are uniquely determined by the equation 
\[
g\xi_i = \xi_{i'}\beta \; ,
\]
and $\Gamma^{(\rho)}$ is the matrix for the representation $\rho$. 

This result follows from a standard construction in induced
representation theory (cf. discussion of the Poincar\'e group in
\cite{representation}). 

The quantum double $D(H)$ and the algebra ${\cal A}^{(1)}$ are 
particular cases of ${\cal A}$. The quantum double is obtained by 
taking $X=H$, $G=H$, with the action $\alpha _g(h)=ghg^{-1}$. As for the
algebra of a single geon, one takes
\[
X=H\times H\,\,\,
\] 
and for the group $G$ the product $H\times {\cal M}$. The actions of 
$\hat\delta _g\in H$ and
$W\in {\cal M}$ commute and are given by 
\[
\alpha_g (a,b) \, =\, 
(gag^{-1} ,gbg^{-1}), \;\,\,\,g\in H 
\]
and 
\[
\alpha_W (a,b) \, =\, (w^{(a)},w^{(b)}), \;\,\,\, W\in {\cal M} ,
\]
where we have used the notation of (\ref{Waction}). 
The IRR's for the algebra (\ref{productinA}) can be
constructed given an element $(a,b) \in H\times H$. The stability
subgroup $K_{(a,b)}\subset H \times H$ is defined by 
\be
K_{(a,b)} \, =\, \left\{ (g,W) \in H\times {\cal M} \mid 
\alpha_g\alpha_W (a,b):=(gw^{(a)}g^{-1},gw^{(b)}g^{-1})=(a,b) \right\} \; . 
\ee
Then, after choosing representatives $\xi_1 , \ldots , \xi_N$ for the 
cosets, the partition of $H\times {\cal M}$ can be written as
\be
\label{K}
H\times {\cal M} \, =\, \xi_1 K_{(a,b)} \cup \xi_2 K_{(a,b)} \cup \cdots \cup 
\xi_N K_{(a,b)} \; .
\ee
Let $\mid 1 \rangle, \ldots , \mid n \rangle \in {\IC} (H\times {\cal
  M})$ be a basis of an IRR of $K_{(a,b)}$.
Then, according to the theorem, the vectors
\be
\label{IREP}
\mid \alpha_{\xi_i}(a, b) \rangle \otimes  \mid j \rangle_{\rho} \;,
\ee
with $i=1\ldots N$, $j=1\ldots n$,
form a basis of an IRR of the algebra 
${\cal A}^{(1)}$. 

Let us express the representations of ${\cal A}^{(1)}$ in a more compact 
notation. 
The action of $H\times {\cal M}$ on $X=H\times H$ divides $X$
into orbits. We denote by $[a,b]$ the orbit containing  the element 
$(a,b)\in H\times H$. We will collectivelly call $\rho $ the quantum numbers  
labeling  the IRR's  of  $K_{(a,b)}$. 
One can see from (\ref{IREP}) that an IRR  $r$ is 
characterized by a pair $r=([a,b],\rho )$. A basis for an IRR $r$ 
of ${\cal A}^{(1)}$ will therefore be written as vectors 
$\mid i,j\rangle ^{(a,b)}_r$ ,$i=1,...,N$; $j=1,...,n$ defined by
\be
\label{IREPvec}
\mid i,j\rangle ^{(a,b)}_r:=\xi_i \mid a,b\rangle\otimes 
\mid j\rangle _\rho   
\ee
where $\mid a,b\rangle $ is a state in the defining representation, 
$\xi_i$ are the same as in (\ref{K}) and $\mid j\rangle _\rho $ are  
base elements in the irreducible representations $\rho $ of
$K_{(a,b)}$. Of course, the set of vectors thus defined depend on the
pair $(a,b)$ we choose. We fix an $a$ and a $b$, and henceforth omit
the superscript. 

The action of $Q_{(a',b')}$ is given by
\be
Q_{(a',b')}|i,j>_r=Q_{(a',b')}~~\xi_i|a,b>\otimes |j>_\rho=
Q_{(a',b')}|a_i,b_i>\otimes |j>_\rho=\delta_{a',a_i}\delta_{b',b_i}|i,j>_r.
\ee

Let $\hat\delta_gW$ be a generic element of  $H\times {\cal M}$. The equation
\be
\label{gaction}
\hat\delta_gW\,\,\xi_i=\xi_{i'}\beta 
\ee
defines uniquely a new class $\xi_{i'}$, together with an element of the 
stability group $\beta \in K_{(a,b)}$. 
The action  of $\hat \delta _gW\in {\cal A}^{(1)}$ on
$\mid i,j\rangle _r$ is determined by (\ref{gaction}) and it reads
\bea
\label{Aaction}
\hat \delta _gW\mid i,j\rangle _r &=& 
\xi_{i'}\mid a,b\rangle\otimes 
\beta \mid j\rangle _\rho =\nonumber \\
&=& \sum_{k}\,{\Gamma ^{(\rho)}(\beta )}_{kj}
\mid i',k\rangle _r
\eea
where $\Gamma ^{(\rho )}$ is the matrix representation of $K_{(a,b)}$.

Each IRR $r=([a,b],\rho )$ describes a distinct 
quantum geon. The corresponding vector spaces ${\cal H}_r$ generated by states
$\mid i,j\rangle _r $, are all finite dimensional. Therefore we can 
easily make it into a Hilbert space by introducing the scalar product
\be
\langle i',j'\mid i,j \rangle _r=\delta _{ii'}\delta _{jj'}.
\ee  

Since the algebras ${\cal A}^{(1)}$ are not the same for different 
choices of the discrete group $H$, we cannot say in general what is
the spectrum of a geon. First, we need to fix a group $H$ and then 
compute the spectrum for the corresponding ${\cal A}^{(1)}$.

Consider now two geons described by representations $r_1$ and $r_2$. The 
associated  Hilbert space of states is simply 
\be
\label{2geonsHilbert}
{\cal H}^{(12)}:={\cal H}_{r_1}\otimes {\cal H}_{r_2}.
\ee
As explained in Section \ref{S3}, the field algebra consists of 
${\cal A}^{(1)}\otimes {\cal A}^{(1)}$ together with ${\cal R}$
and  ${\cal S}$. The elements of ${\cal A}^{(1)}\otimes {\cal A}^{(1)}$ act
naturally on (\ref{2geonsHilbert}). It remains to be said what is the
action of  ${\cal R}$ and ${\cal S}$ on states in ${\cal
  H}_{r_1}\otimes {\cal H}_{r_2}$.  

The action of ${\cal R}$ is completely determined by the formula
(\ref{R}):
\[
{\cal R}=\sigma \sum _{a,b}Q_{(a,b)}\otimes {\hat \delta}^{-1} _{aba^{-1}b^{-1}}.
\]
In other words
\be
\label{actR}
{\cal R}\mid i,j \rangle _{r_1}\otimes \mid k,l\rangle _{r_2}=
\sum _{a,b}{\hat \delta}^{-1}_{aba^{-1}b^{-1}}\mid k,l \rangle _{r_2}\otimes 
Q_{(a,b)} \mid i,j\rangle _{r_1}.
\ee

The generalization for $n$ geons is straightforward.

We may think of ${\cal R}$ and ${\cal S}$ as scattering matrices for a
pair of geons. The ${\cal R}$-matrix represents an ``elastic'' interaction in
the sense that two incoming geons of quantum numbers $r_1$ and $r_2$
are scattered into two objects carrying the same quantum numbers 
$r_1$ and $r_2$. The
handle slide ${\cal S}$ on the contrary is a nontrivial
scattering, each one of the two outgoing geons being a
superposition of many geons in the spectrum.

\sxn{Quantum Topology Change}\label{S5}

In this paper we have considered $(2+1)d$ manifolds such that any spatial slice
consists of a plane with a certain number $n$ of handles. In other words,
for each fixed time, the configuration consists of $n$ geons. 
If the number of geons is not fixed, 
we say that topology can change. Creation of baby universes is also 
a topology-changing process, but we will not consider it here for reasons that 
should become clear in what follows. 

Our system is described by a certain field algebra ${\cal A}^{(n)}$,
and its quantization is given by a representation of ${\cal A}^{(n)}$. A 
change in the number of geons means necessarily a change in the field
algebra. Let us see how that can be accomplished. Let us suppose that 
a geon, represented by a square with opposite sides identified, has a typical
size $l$ that can vary with time. 
Intuitively, a geon can disappear if $l$ becomes too 
small. In this case, a geon will resemble a point-like object. 
Let us consider  the limiting case $l\rt 0$. It is clear that the holonomies 
associated with loops $\gamma _1$ and $\gamma _2$ of Fig. 3.1 do not 
make sense in this limit. The only flux observable available in this
limit is the  
holonomy of  $\gamma _3$, responsible for measuring the  total flux. 
The 
algebra describing the limiting situation is the one-particle approximation 
${\cal A}_L^{(1)}\subset {\cal A}^{(1)}$ introduced in Section 
\ref{GASP}. Actually,
we do not need to consider the limit $l\rt 0$, since our description is 
supposed to be an effective theory that is not valid beyond a certain 
scale of energy (distance). 
We may say that after the geon has become very small, 
the operators associated with individual fluxes no longer belong to the 
low energy (large distance) description. 

The structure of the field algebra tells us that a geon can turn into
a point-like object, but it cannot disappear. However, this is only a
semi-classical description. 

The quantum theory is described by states belonging to 
an IRR $r$ of ${\cal A}^{(1)}$. From the inclusion
\be
i^{(1)}:{\cal A}_L^{(1)}\hookrightarrow  {\cal A}^{(1)},
\ee
it follows that $r$ is also a (in general reducible) representation of ${\cal
  A}_L^{(1)}$. Let ${\cal H}_r$ be the vector space carrying the
representation $r$. In general,
${\cal H}_r$ is decomposable as a direct sum 
\be
\label{decomp}
{\cal H}_r=\bigoplus _\sigma N^r_\sigma V_\sigma ,\,\,\,\, N^r_\sigma \in \IN 
\ee 
where $V_\sigma $ carries the IRR $\sigma$ of ${\cal A}^{(1)}_L$. The
long distance observer does not see operators mixing different IRR's. 

Therefore, a long distance observer interprets (\ref{decomp}) as saying
that a geon carrying a representation  $r$ can decay into different
{\em particles} carrying representations $\sigma $. It could happen
that the trivial representation $\sigma =0$ of ${\cal A}^{(1)}_L$
occurs in (\ref{decomp}). In this case, for an observer working only
with ${\cal A}^{(1)}_L$, {\em there will be a non-zero probability of
seeing the vacuum}.

As an example, let us next characterize the vacuum representation and
discuss vacuum decay.

The IRR's of ${\cal A}^{(1)}_L$ 
are classified in a similar way as for 
${\cal A}^{(1)}$. The trivial IRR on $V_0$ in the decomposition (\ref{decomp})
is generated by any state $\mid \mbox{VAC}\rangle \in {\cal H}_r$ 
satisfying
\bea
\label{VAC}
Q^{(1)}_c\mid \mbox{VAC}\rangle 
&=& \delta _{e,c}\, \mid \mbox{VAC}\rangle \,,\\
\hat \delta ^{(1)}_g \mid \mbox{VAC}\rangle 
&=& \mid \mbox{VAC}\rangle \, , \label{v2}\\
C^{(1)} \mid \mbox{VAC}\rangle &=& \mid \mbox{VAC}\rangle\,.
\label{v3} 
\eea

We will call such state vacuum. It is not difficult to show that a representation $r=([a,b],\rho )$ 
contains states satisfying (\ref{VAC}) if and only if  
$aba^{-1}b^{-1}$ is the identity. Furthermore, under the condition 
\[
aba^{-1}b^{-1}=e,
\] 
all states of ${\cal H}_r$,  $r=([a,b],\rho )$, fulfill equation (\ref{VAC}). 
We are thus left with the conditions (\ref{v2}) and 
(\ref{v3}) for defining the vacuum state. They simply mean that 
$\mid VAC \rangle $ is an identity representation  of the group 
$H\times \IZ_N$, where $\IZ_N$ is generated by $C^{(1)}=\hat C_{2\pi }$. 

Note that vacuum decay occurs naturally, for example, in all IRR's
of ${\cal A}^{(1)}$ of the form $r=([a,b],\epsilon )$, where $a$ and
$b$ are in the center of $H$ and $\epsilon$ is the trivial
representation of the stability subgroup of $(a,b)$, which in this
case is the whole of $H \times {\cal M}$. The vectors in this
representation clearly satisfy all conditions and therefore will decay
into vacuum states.

The vector space
${\cal H}_r$ may contain more than one copy of the identity representation
of  $H\times \IZ_N$. We will denote the set of corresponding
orthonormal vectors by 
\[
\mid \mbox{VAC};l\rangle ,\,\,\,\,l=1,2,...,N_0^r.
\]
Finding all $\mid VAC;l\rangle $ in a given decomposition of each ${\cal
H}_r$ is a group theoretical problem that can 
be solved for specific choices of the discrete group $H$. We shall not
attempt this here.

The probability $P(\psi )$ of a normalized state 
$\mid \psi \rangle \in {\cal H}_r $ to  decay into the vacuum is then given by
\be
\label{prob}
P(\psi )= \left\{ \begin{array}{ll}
        \sum _{l} 
       |\langle \mbox{VAC};l\mid \psi \rangle|^2 \neq 0 \ & \mbox{ if }N_0^r\neq 0,\\
      & \\       
    =0 &\mbox{ if }N_0^r=0.
\end{array}
         \right.
\ee

If $N_0^r= 0$, a single geon described by $r$ cannot decay 
into the vacuum. However, two geons colored by $r$ and  $r'$ may annihilate
each other. The two geons can shrink to localized objects and at the same
time approach each other. The process can also be interpreted as a
change in the scale of observations to long distances. 
At the end a point-like object will remain
and should be described by the algebra  
${\cal  A}^{(2)}_L$ introduced in Section \ref{GASP}. From the inclusion
\be
i^{(2)}:{\cal A}^{(2)}_L\hookrightarrow {\cal A}^{(2)}
\ee
follows that the space of states ${\cal H}_r\otimes {\cal H}_{r'}$ of
the two geons is a (reducible) representation of ${\cal
  A}^{(2)}_L$. Let $\sigma $ denote as before the IRR's 
of ${\cal A}^{(2)}_L$, with corresponding vector spaces $V_\sigma$. Then
\be
{\cal H}_r\otimes {\cal H}_{r'}=
\bigoplus _\sigma N^{(r,r')}_\sigma V_\sigma ,\,\,\,\, N^{(r,r')}_\sigma \in \IN. 
\ee 
Therefore, the operators of  ${\cal A}^{(2)}_L$ can see the vacuum
if $N^{(r,r')}_0$ is not zero. The vacuum representation and the vacuum
probability decay are given by formulae analogous to
(\ref{VAC})-(\ref{prob}).

It is clear now how to describe the decay into the vacuum of an
arbitrary number of geons. Consider $n$ geons described by
representations $r_1,...,r_n$. The space of states 
${\cal H}_{r_1}\otimes ...\otimes {\cal H}_{r_2}$ is a representation of
the long distance algebra ${\cal A}^{(n)}_L\subset {\cal A}^{(n)}$
described in Section \ref{GASP}. The system may decay into the
vacuum if  this representation contains the trivial  representation 
of ${\cal A}^{(n)}_L$. 

\sxn{Concluding Remarks}\label{Final}

In this work we have developed an algebraic model for topological geons which
describes topology change as a purely quantum phenomenon rather than
the usual classical sense of
cobordisms between two non-homeomorphic spatial manifolds $\Sigma$ and
$\Sigma'$. Instead, our formalism revealed what an observer, probing
the topology of space by using only quantum operators and quantum states,
would be able to see. 

The key point was that in resorting to a field theory to infer the
underlying spatial topology, one could only take into account those
operators which were compatible with the scale of observations, since
no other operator would a have sensible physical meaning in the
theory. The passage from a larger scale of observations to a smaller one was
represented, on a more technical level, by selecting a subalgebra of
the original algebra describing the system in the quantum theory. The
quantum states of the system, which in the case of geons give a direct
information on the spatial topology, could now decay into the vacuum,
leading the would-be observer to conclude that a topology change has
ocurred. 

There is another, perhaps more intuitive view of the sort of topology
change we have envisaged in this paper. As pointed out in Section
\ref{S3} for the case of vortices, those classical configurations for
which holonomies are trivial around some ``topological blob'',
be it a vortex or a geon, are indistinguishable from those in which
this ``blob'' is absent, or ``vacuum'' configurations. If we view
quantum states as wave functions, it is clear that their role is to
assign a probability to each classical configuration. A quantum
transition to states which are very sharply peaked, or localized at
the aforementioned ``vacuum'' configurations will be interpreted by an
observer as a quantum topology change. Such states
correspond to the vacuum states of Section \ref{S5}. 

We have restricted ourselves to a simple theory, where complications
arising from local degrees of freedom were absent (the theory we
considered is topological in the limit of very low energies), and we
could concentrate on the topological aspects more unobstrusively.

Of independent value is the algebra describing the topological
geons. The finite group $H$ can be generalized to a Lie group $G$. Of
special interest is the case $G = SO(2,1)$, which
describes geons in the presence of gravity. A suitable generalization
of our formalism promptly discloses a whole spectrum of geon types in
quantum gravity, and many interesting properties of these entities can be
explored, as for instance spin-statistics connection. Although this
issue has been extensively studied in the literature, our formalism
may shed new light on some points. This subject will be investigated
in a forthcoming paper \cite{us}.     

\newpage
\vspace{.5cm}
\noindent
{\large \bf Acknowledgments}

\noindent
We would like to thank J.C.A. Barata, B.G.C. da Cunha, A. Momen and S. Vaidya
for many helpful discussions. The work of A.P. Balachandran was
supported by the Department of
Energy, U.S.A., under contract number DE-FG02-85ERR40231. The work of
E. Batista, I.P. Costa e Silva and
P. Teotonio-Sobrinho was supported by FAPESP, CAPES and CNPq
respectively. A.P.B. acknowledges the wonderful hospitality of Denjoe
O'Connor and support of CINVESTAV at Mexico, D.F. while this work was
being completed.

\end{document}